\documentclass[useAMS,usenatbib]{mn2e}
\usepackage{amsmath}
\usepackage{amssymb}
\usepackage[dvips]{graphicx}
\usepackage{verbatim}

\usepackage{color}

\def \BE{\begin{equation}}
\def \EE{\end{equation}}

\def \HI{ h^{-1}}
\def \MSUN{M_{\odot}}
\def \NPART{\rm N_{\rm part}}
\def \LBOX{\rm L_{\rm box}}
\def \MPART{M_{\rm part}}
\def \LRES{\rm L_{\rm soft}}

\def \FNL{f_{\rm NL}}
\def \SIGMA8{\sigma_{8}}
\def \OM{\Omega_{\rm m}}
\def \OB{\Omega_{\rm b}}
\def \OL{\Omega_{\rm \Lambda}}

\def \SG{\sigma}
\def \SGI{\sigma^{-1}}

\def \DELTAC{\delta_{\rm c}}

\usepackage{color}
\definecolor{orange}{rgb}{1,0.5,0}

\title[Mass function and bias from non-Gaussian simulations]{Halo mass function
and scale-dependent bias from N-body simulations with 
non-Gaussian initial conditions}
\author[Pillepich, Porciani \& Hahn]{Annalisa Pillepich$^{1}$\thanks{E-mail:
annalisa@phys.ethz.ch}, Cristiano Porciani$^{1,2}$ and Oliver Hahn$^{1}$\\
$^{1}$Institute for Astronomy, ETH Zurich, 8093 Zurich, 
Switzerland\\
$^{2}$Argelander Institut f\"ur Astronomie, Auf dem H\"ugel 71, D-53121,
Germany}
\begin{document}

\maketitle
\label{firstpage}


\begin{abstract}
We perform a series of high-resolution N-body simulations of cosmological
structure formation
starting from Gaussian and non-Gaussian initial conditions. 
We adopt the best-fitting cosmological parameters from the third- and fifth-year
data releases of the {\sl Wilkinson Microwave Anisotropy Probe} and
we consider non-Gaussianity of the local type parameterised by
eight different values of the non-linearity parameter $\FNL$. 
Building upon previous work based on the Gaussian case,
we show that,  when expressed in terms of suitable variables,
the mass function of friends-of-friends haloes
is approximately universal (i.e. independent of redshift, cosmology,
and matter transfer function) to good precision (nearly 10 per cent)
also in non-Gaussian scenarios.
We provide fitting formulae for the high-mass end 
($M>10^{13} \,\HI M_\odot$) of the
universal mass function in terms of $\FNL$, and we also present  
a non-universal fit in terms of both $\FNL$ and $z$ to be used
for applications requiring higher accuracy.
For Gaussian initial conditions, we extend our fit to a wider range of
halo masses ($M>2.4 \times 10^{10}\, \HI M_\odot$) and we also provide
a consistent fit of the linear halo bias.
We show that, for realistic values of $\FNL$, 
the matter power-spectrum in non-Gaussian cosmologies departs
from the Gaussian one by up to two per cent on the scales where
the baryonic-oscillation features are imprinted on the two-point statistics. 
Finally, using both the halo power spectrum and the halo-matter cross spectrum, 
we confirm the strong $k$-dependence of the halo bias on large scales
($k<0.05\, h$ Mpc$^{-1}$) which was already detected in previous studies. 
However, we find that commonly used
parameterisations based on the peak-background split do not provide an
accurate description of our simulations which present extra dependencies
on the wavenumber, the non-linearity parameter and, possibly, 
the clustering strength.
We provide an accurate fit of the simulation data that can be used
as a benchmark for future determinations of $\FNL$
with galaxy surveys.
\end{abstract}

\begin{keywords}
cosmology: theory, dark matter, large-scale structure -- methods: N-body
simulations -- galaxies: haloes, clusters.
\end{keywords}


\section{Introduction}
The detection of temperature anisotropies in the Cosmic Microwave Background
(CMB) provided evidence 
that large-scale structure formation in the universe
was seeded by small density fluctuations generated at early times.
The statistical properties of these seeds are usually modelled with a Gaussian 
random field.
Historically the Gaussian approximation was introduced for mathematical 
convenience. In the absence of a solid model for the generation of
density fluctuations the Gaussian hypothesis was accepted on the basis
of the central limit theorem 
(e.g. \citealt{BARD} and references therein).
The advent of inflationary models provided further support for Gaussianity.
Small-amplitude curvature perturbations generated during a standard 
inflationary phase (single field, slow roll) are very nearly Gaussian 
distributed (e.g. \citealt{BKMR} and references therein).

However, many variants of the inflationary scenario predict 
appreciable levels of primordial non-Gaussianity.
In terms of Bardeen's gauge-invariant potential, $\Phi$, most of these 
models \citep[but not all, see e.g.][]{CREM} can be reduced to the
form:
\BE
\Phi = \phi +\FNL (\phi^2-\langle \phi^2\rangle)\;,
\label{deffnl}
\EE
where $\phi$ is an auxiliary Gaussian random field and $\FNL$ quantifies the 
amount of primordial non-Gaussianity. On subhorizon scales, $\Phi=-\Psi$ where 
$\Psi$ denotes the
usual peculiar gravitational potential related to density fluctuations via
Poisson's equation. The parameter $\FNL$ thus has the same sign as the skewness
of the density probability distribution function.
This local form of non-Gaussianity (note that equation (\ref{deffnl}) applies 
in configuration space) can be obtained from a truncated expansion
of the effective inflaton potential \citep{SB,FRS,GANGUI}. The parameter $\FNL$
thus encodes
information about the inflaton physics.
Standard inflation gives $|\FNL |\ll 1$ \citep{SB,MALD}.
However, even in 
this case, the non-linear evolution of 
perturbations on superhorizon scales yields an observable $\FNL$ of 
order unity (which, in reality, should be scale and redshift dependent;
\citealt{BMR}, see also \citealt{PC}). 
Large values of $|\FNL|$ naturally arise in multi-field inflation models
(e.g. \citealt{LM}; for an extensive review see \citealt{BKMR}) and even in
cyclic or ekpyrotic 
models of the universe with no inflation \citep{CS,BKO,LS}.

Observational constraints on $\FNL$ have been derived studying
three-point statistics of temperature fluctuations in the CMB \citep{KS}. 
The recent 5-year data from
the {\sl Wilkinson Microwave Anisotropy Probe} (WMAP) give
$-9<\FNL<111$ at the 95 per cent confidence level \citep{KOMATSU}. 
Parallel studies on the same dataset give 
$-178<\FNL<64$ using Minkowski functionals (Komatsu et al. 2008) 
and $-8<\FNL <111$ from wavelet decomposition \citep{CURTOETAL}.
Some recent reanalyses of earlier 3-year WMAP data claim substantial evidence
for positive $\FNL$:
$27<\FNL<147$ from the bispectrum of temperature fluctuations 
\citep{YW}
and $23<\FNL<75$ from their 
one-point distribution function \citep{JS}.
On the other hand, a study of Minkowski functionals on the 3-year data gives
$-70<\FNL<91$ \citep{HIKAGE}.
Higher quality data are needed to improve these constraints.
The upcoming Planck satellite should be able to reduce the uncertainty in 
$\FNL$ to $\sim 5$ \citep{KS}. 

Alternatively, one might use observational
signatures of primordial non-Gaussianity imprinted in the large-scale
structure (LSS) of the universe (e.g. \citealt{MOSCA}).
Ideally, one would like to use high-redshift probes as the
non-linear growth of density fluctuations quickly superimposes a strong 
non-Gaussian signal onto the primordial one 
so that the latter might then be difficult to recover.
For instance, the large-scale distribution of neutral 
hydrogen in the era between hydrogen recombination and reionisation encodes
information on $\FNL$ \citep{PPM}. This could
be probed by detecting the redshifted hyperfine 21-cm transition with 
very low-frequency radio arrays from space. In principle, an experiment
of this kind can limit $\FNL$ to $\Delta \FNL<1$ 
(\citealt{PPM}, see also \citealt{COO}).  
However, it is not clear yet whether such an experiment will ever be possible
due to technical complexity and problematic foreground subtraction.
At lower redshifts, $\FNL$ can be constrained probing
the statistics of rare events, as like as
the mass function of galaxy groups and clusters
\citep{MLB,MVJ,KOYAMA,RB,RGS,LOVERDE}.
Early attempts of using cluster counts to constrain $\FNL$ have been
rather inconclusive due to low-number statistics
(see e.g. \citealt{WILL,AR} and references therein).
Even though cluster-mass estimates are still rather uncertain and 
massive objects are very rare, the observational perspectives
look very promising. A number of galaxy surveys encompassing large fractions
of the observable universe
are being planned (e.g. ground-based surveys as DES, PanSTARRS, and LSST, 
and the satellite missions EUCLID and ADEPT) and could
potentially lead to solid measurements of $\FNL$ \citep[e.g.][]{DALAL,CVM}.

\begin{table*}
 \centering
  \caption{Specifics of the N-body simulations.}
  \label{TAB_RUNS}	
  \begin{tabular}{|cccccccc|}
  \hline
  Name	& $\FNL$& $\NPART$	& $\LBOX$	& $\MPART$	& $\LRES$
& $z_{\rm start}$ & Cosmology  \\
	&	&	& $(\HI {\rm Mpc})$	& $(\HI \MSUN )$& $(\HI
{\rm kpc})$	& 		&	\\
 \hline
 1.750& +750& $1024^3$ & 1200 & $1.246 \times 10^{11}$ & 20 & 50 &WMAP5 \\
 1.500 &+500& $1024^3$ & 1200 & $1.246 \times 10^{11}$ & 20 & 50 &WMAP5\\
 1.250 & +250& $1024^3$ & 1200 & $1.246 \times 10^{11}$ & 20 & 50 &WMAP5 \\
 1.80 & +80& $1024^3$ & 1200 & $1.246 \times 10^{11}$ & 20 & 50 &WMAP5 \\
 1.27 & +27& $1024^3$ & 1200 & $1.246 \times 10^{11}$ & 20 & 50 &WMAP5 \\
 1.0 &  0 & $1024^3$ & 1200 & $1.246 \times 10^{11}$ & 20 & 50 & WMAP5 \\
 1.-27 & -27& $1024^3$ & 1200 & $1.246 \times 10^{11}$ & 20 & 50 &WMAP5 \\
 1.-80 &-80& $1024^3$ & 1200 & $1.246 \times 10^{11}$ & 20 & 50 &WMAP5\\
\hline
 2.0 & 0 & $1024^3$ & 1200 & $1.072 \times 10^{11}$ & 20 & 50 &WMAP3 \\
 2.750 & +750& $1024^3$ & 1200 & $1.072 \times 10^{11}$ & 20 & 50 & WMAP3\\
 \hline
 3.0 & 0 & $1024^3$ & 150 & $2.433 \times 10^{8}$ & 3 & 70 &WMAP5\\
 3.250 & +250& $1024^3$ & 150 & $2.433 \times 10^{8}$ & 3 & 70 &WMAP5 \\
 \hline
\end{tabular}
\end{table*}	

Primordial non-Gaussianity is also expected to modify the clustering
properties of massive cosmic structures forming out of rare density
fluctuations \citep{GW,MLB,LMV,KOYAMA}. 
Also in this case, however,
the non-linear evolution of the mass density generally
superimposes a stronger signal than that generated by primordial 
non-Gaussianity onto the galaxy three-point statistics.
The galaxy bispectrum is thus sensitive to $\FNL$ only at high redshift 
\citep{VWHK,SSZ,SK}.

Recently, \citet{DALAL} have shown analytically
that primordial non-Gaussianity
of the local type is expected to generate a scale-dependent large-scale 
bias in the clustering properties of massive dark-matter haloes.
This is a consequence of the fact that large and small-scale density
fluctuations are not independent when $\FNL\neq 0$.
Similar calculations have been presented by \citet{MV}, \citet{SLOSAR}, 
\citet{AT}, and \citet{DONALD}. 
Numerical simulations by \citet{DALAL} are in qualitative agreement
with the analytical predictions confirming the presence of a scale-dependent
bias. 
Using these analytical models for halo biasing to describe the clustering
amplitude of luminous red galaxies and quasars 
from the {\sl Sloan Digital Sky Survey}, 
\citet{SLOSAR} obtained $-29<\FNL<69$ at the 95 per cent confidence level.
This shows that LSS studies are competitive with CMB experiments to constrain
primordial non-Gaussianity but also calls for more accurate 
parameterisations of the mass function and clustering statistics
of dark-matter haloes arising from non-Gaussian initial conditions.

Most of the analytic derivations of the non-Gaussian halo mass function
\citep[e.g.]{MVJ,LOVERDE}
are based on the extended Press-Schechter model \citep{PS,BOND} 
which, in the Gaussian case, is known to produce inaccurate 
estimates of halo abundance \citep{ST,JENK}.
Similarly, the scale dependent bias is obtained either 
using the peak-background split model \citep{SLOSAR} 
or assuming that haloes form from
the highest linear density peaks \citep{MV}. 
Both techniques have limited validity
in the Gaussian case \citep{JING,PCL,ST}.
In this paper we test the accuracy of the excursion-set model and
the peak-background split in the non-Gaussian case. 
This extends the previous studies of \citet{KANG}, \citet{GROSSI1} and
\citet{DALAL}
for the halo mass 
function and of \citet{DALAL} for the halo bias by exploring more
realistic values for $\FNL$ with simulations of better quality.
In practice,
we run a series of high-resolution N-body simulations where we follow
the process of structure formation starting from Gaussian and non-Gaussian
initial conditions. The halo mass function and bias extracted from the
simulations are 
then compared with the existing analytical models and 
used to build accurate fitting formulae. 
These will provide a benchmark for future determinations of non-Gaussianity
with galaxy surveys.

The paper is organised as follows.
In Section 2 we describe our N-body simulations. 
In Sections 3, 4, and 5 we present our results for the halo mass function,
the matter power spectrum and the halo bias, respectively.
In Section 6 we discuss the implications of our results for the analysis
by \citet{SLOSAR}. 
Our conclusions are summarised in Section 7.

\begin{table}
 \centering
  \caption{Assumed cosmological parameters.}
  \label{TAB_COSMOL}	
  \begin{tabular}{|ccccccc|}
  \hline
  Name	& $h$& $\sigma_8$ &$n_{\rm s}$& $\OM$	& $\OB$	& $\OL$	  \\
  \hline
  WMAP3& 0.73	& 0.76	& 0.95	& 0.24	& 0.042	 & 0.76 \\	
  WMAP5& 0.701	& 0.817	& 0.96	& 0.279 & 0.0462 & 0.721\\
\hline
\end{tabular}
\end{table}

\section[]{N-body simulations}
\subsection{Specifics of the simulations}
We use the lean version of the tree-PM code {\sc Gadget-2} \citep{SPRINGEL}
kindly made available by Volker Springel
to follow the formation of cosmic structure in a flat
$\Lambda\textrm{CDM}$ cosmology.
We run three different series of simulations (each containing
$1024^3$ collisionless particles) that differ in the adopted
cosmology, box size (and thus force softening length, $L_{\rm soft}$), and 
initial redshift (details are summarised in Table \ref{TAB_RUNS}).
The assumed cosmological parameters are listed in
Table \ref{TAB_COSMOL}.
For our series $\# 1$ and $\# 3$ they coincide with the 
5-yr WMAP best estimates \citep{KOMATSU}.
The combined 3-yr WMAP+LSS results by \citet{SPERGEL} are 
instead used for series $\# 2$.

We produce non-Gaussian initial conditions directly applying
equation (\ref{deffnl}) after having generated the Gaussian random
field $\phi$ with standard Fourier techniques.
We consider eight values for the parameter $\FNL$: 
$-80,-27, 0, +27, +80, +250, +500, +750$.
The first five are within the current constraints from CMB data
\citep{KOMATSU}, while the three largest values are useful to compare
with previous work.
Within each series of simulations, 
we use the same set of random phases to generate
the Gaussian potential $\phi$. This facilitates the comparison
between different runs by minimising sample variance.

The linear matter transfer function, $T(k)$, is computed using the 
{\sc Linger}
code \citep{BERTSCH} and is applied after creating the non-Gaussian
potential $\Phi$ in equation (\ref{deffnl}).
Particle displacements and velocities at $z_{\rm start}$ 
are generated using
the Zel'dovich approximation \citep{ZEL}. A critical discussion of this
choice is presented in the Appendix.

Particle positions and velocities are saved for 30 time steps
logarithmically spaced in $(1+z)^{-1}$ between $z=10$ and $z=0$.
Dark-matter haloes are identified using the standard friends-of-friends
 (FOF)
algorithm with a linking length equal to 0.2 times the mean interparticle
distance. We only considered haloes containing at least 100 particles.

Our first two series of simulations only include large periodic
boxes covering a volume of $(1200 ~ \HI {\rm Mpc})^3$ where we  
can study haloes with masses ranging from $10^{13}$ up to $10^{15}~\HI \MSUN$.
These simulations will be used to analyse both the mass function and the 
bias of dark-matter haloes.
On the other hand, the third series includes simulations covering a volume
of $(150 ~ \HI {\rm Mpc})^3$. 
They will be used to study the
mass function and the 
bias of low-mass haloes with $10^{10}<M<10^{13}\, h^{-1} M_\odot$.

\subsection{A note on the definition of $\FNL$}
The definition of $\FNL$  given in equation (\ref{deffnl}) depends on
the cosmic epoch at which it is applied.
The reason for this time dependence is that both potentials $\Phi$ and 
$\phi$ decay with time  proportionally
to $g(a)=D(a)/a$ with $D(a)$ the linear growth factor of density fluctuations
and $a$ the Robertson-Walker scale factor.

In this paper, we define $\FNL$ by applying equation (\ref{deffnl}) at
early times, namely at $z=\infty$. Other authors have adopted different
conventions. \citet{GROSSI1} use the linearly-extrapolated fields 
at $z=0$ to define $\FNL$. Therefore, their values of the $\FNL$ 
parameter need to be divided by the factor $g(\infty)/g(0)$ to match ours.
In the WMAP5 cosmology, $g(\infty)/g(0)\simeq 1.3064$.
On the other hand, \citet{DALAL} apply
equation (\ref{deffnl}) at $z_{\rm start}$, the redshift at which
they generate the initial conditions for the simulations. This agrees
with our definition to better than 0.01 per cent.

The sign convention for the non-linearity parameter might possibly generate
further ambiguity.
In our simulations,
positive values $\FNL$ correspond to positive skewness of the mass-density
probability distribution function.
The same convention has been adopted by \citet{GROSSI1}, \citet{KANG} and
\citet{DALAL}.

\begin{table*}
  \caption{Widely used parameterisations for the halo mass function
deriving from Gaussian initial conditions. 
}

  \begin{tabular}{c|c|c|c|}
  \hline
  Acronym & Reference & Functional form & Parameters \\
  \hline
PS &  \citet{PS}    	& 
  $f_{\rm PS}(\SG) = \sqrt{\frac{2}{\pi}}~ \frac{\delta_c}{\SG}~ \exp
  \left(-\frac{\delta^2_c}{2\SG^2} \right)
  $     & $\delta_c=1.686$     \\	
ST &  \citet{ST}  	& $
f_{\rm ST}(\SG) = A \sqrt{\frac{2 a}{\pi}}~ \frac{\delta_c}{\SG} \exp
\left(-\frac{a ~\delta^2_c}{2\SG^2} \right)~\left[1+\left(
\frac{\SG^2}{a ~\delta^2_c}\right)^p \right]
$   &  $A=0.322, a=0.707, p=0.3$ \\
J &  \citet{JENK}		& 
$
f_{\rm J}(\SG) = A \exp{\left(-\vert\rm{ln}\, \SG^{-1} + B \vert^{\rm p}
\right)}
$  & $A=0.315, B=0.61, p=3.8$ \\
R &  \citet{REED}		& $
f_{\rm R}(\SG) = f_{\rm ST}(\SG) \exp{\left( \frac{-a}{\SG
(\cosh{2\SG})^{b}}\right)}
$    & $a=0.7, b=5$ \\
W &  \citet{WARR}		& $
f_{\rm W}(\SG) = A \left(\SG^{-a}+b \right)
\exp{\left(-\frac{c}{\SG^2} \right)}
$  &A=0.7234, a=1.625, b=0.2538, c=1.1982 \\
T &  \citet{TINK}	&$
f_{\rm T}(\SG) = A\left[ \left(\frac{\SG}{b}
\right)^{-{a}}+1\right]\exp{\left(-\frac{c}{\SG^2} \right)}
$ & vary with halo overdensity\\
\hline
\end{tabular}
  \label{TAB_MF}
\end{table*}

\begin{figure}
  \includegraphics[width=8cm]{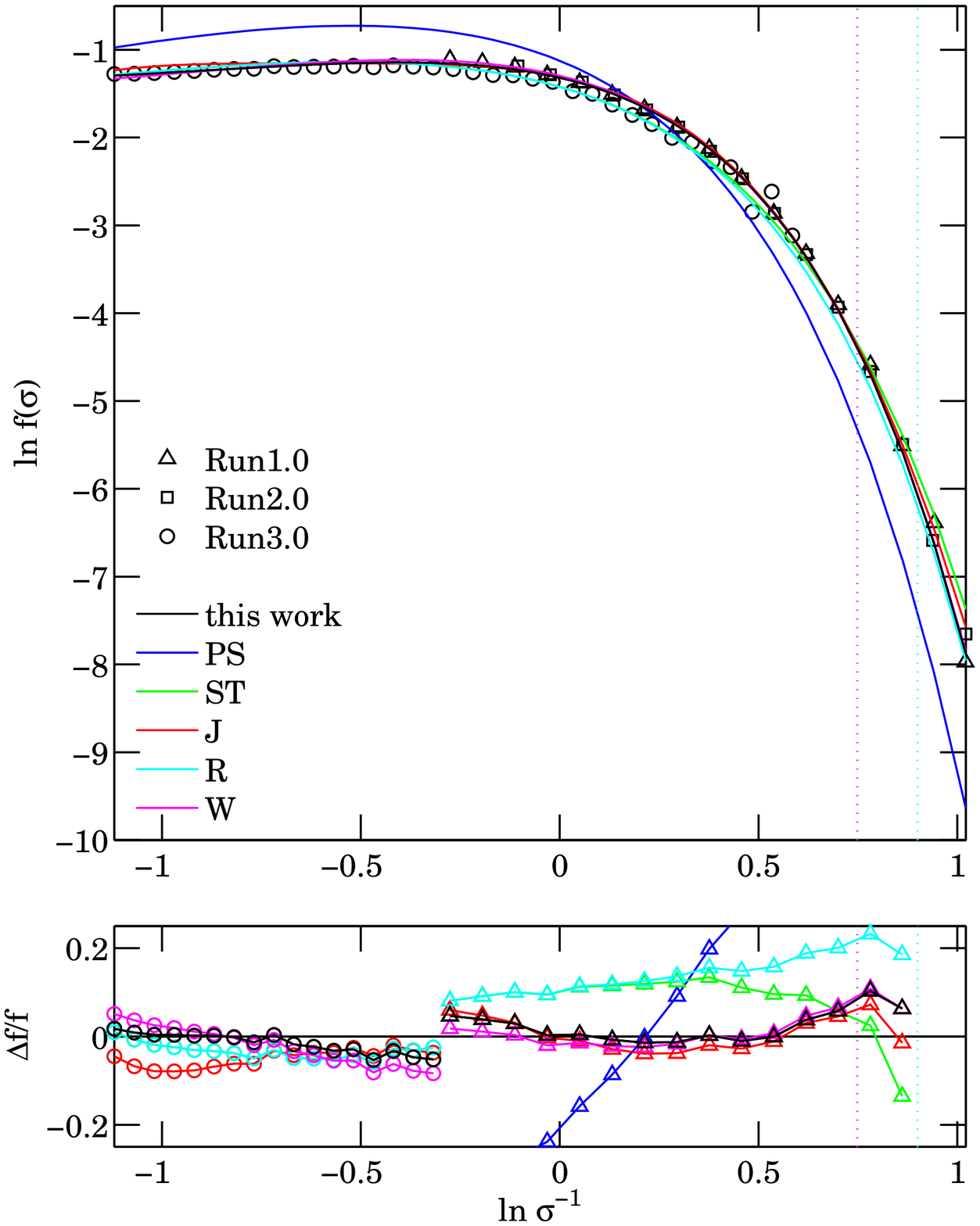}
  \caption{The universal mass function in our Gaussian simulations
Run 1.0 (triangles), Run 2.0 (squares), and Run 3.0 (circles)
is compared with a number of fitting formulae listed in Table \ref{TAB_MF}.
Data are equispaced in ${\rm ln}\, \SGI$ and only bins containing more than
30 haloes are shown. 
The vertical dotted lines indicate the upper mass limits used in
\citet{JENK}, \citet{REED}, and \citet{WARR}. 
The corresponding
low-mass limits are all equal or smaller than $ \rm{ln}\,\SGI = -1.2$.
The lower panel shows
the residuals $\frac{\Delta f}{f} = \frac{({\rm data} - {\rm fit})}{\rm data}$ 
between
our data points and the different fitting functions. 
Here we only show data with a Poisson uncertainty better than 5 per cent.
For clarity only outputs from Run 1.0 (triangles, $ \rm{ln}\,\SGI > -0.3$) 
and Run 3.0 (circles, $ \rm{ln}\,\SGI < -0.3$ )
are plotted.}
  \label{FIG_GAUSSMF}
\end{figure}

\section[]{The Halo Mass Function}
One of the long standing efforts in cosmology is to determine
the mass function of dark matter haloes $dn/dM(M,z)$ -- 
i.e. the number of haloes 
per unit volume per unit mass at redshift $z$ -- 
from the statistical properties of the linear density field.
Analytical work has suggested that, when expressed in terms of suitable
variables,
the functional form of $dn/dM$ should be universal to changes in
redshift and cosmology 
\citep{PS,BOND,ST}.
N-body simulations have shown that this is approximately true
when structure formation is seeded by Gaussian perturbations
\citep{JENK,EVRARD,WHITE,WARR,TINK}.

Following these studies, we describe the halo abundance in our simulations
through the following functional form
\BE
\frac{dn}{dM}(M,z) = f(\SG)~ \frac{\bar{\rho}_{\rm {m}}}{M}\frac{{d}
~{\rm ln}
[\SG^{-1}(M,z)]}{{d}M}\;.
\label{EQ_MF}
\EE
where $\bar{\rho}_{\rm{
m}}$ is the mean background matter density today, 
and $\SG^2(M,z)$ is the variance of the linear density field 
\BE
\SG^2(M,z) = \frac{1}{2 \pi^2} \int^{\infty}_{0} k^2 ~P_{\rm lin}(k,z)~
W^2(k,M)\, {d}k,
\EE
with $P_{\rm lin}(k,z)$ the corresponding power spectrum and
$W^2(k,M)$ some window function with mass resolution $M$
(here top-hat in real space).
The validity of equation (\ref{EQ_MF}) has been widely tested against
numerical simulations and useful parameterisations for $f(\SG)$ have been
provided starting from Gaussian initial conditions
\citep{ST,JENK,WARR}.
These fitting functions have an accuracy ranging from 5 to 20 per cent
depending on redshift, cosmology, and the exact definition of halo masses.
Recently, \citet{TINK} have detected deviations from universality
in $f(\SG)$: redshift-dependent corrections are needed to match 
the mass function in simulations with an accuracy of 5 per cent.
This result is based on
haloes identified with the spherical overdensity
algorithm. It is well known that the mass function of FOF haloes shows
a more universal scaling even though other halo finders might be more
directly linked to actual observables \citep{JENK,TINK}. 
Deviations from universality for FOF haloes
will be further discussed in Section 3.3.
One should anyway keep in mind that baryonic physics can cause 30 per cent
deviations in $dn/dM$ with respect to the pure dark-matter case 
\citep{SRE}.

\subsection{Halo mass function from Gaussian initial conditions}

The halo mass functions extracted 
from our Gaussian simulations -- Run 1.0 (triangles), Run 2.0 (squares), and
Run 3.0 (circles) -- are presented in Figure \ref{FIG_GAUSSMF}.
The combination of different box sizes allows us to cover the very wide
range $ -1.2 < \rm{ln}\,\SGI < 1.1$ which
roughly corresponds to the mass interval $2 \times 10^{10} < M < 5
\times 10^{15} \,\HI \MSUN $ at $z=0$.
Figure \ref{FIG_GAUSSMF} has been obtained by combining data from snapshots
at redshifts $z<1.6$.
Note that, at a fixed redshift, 
larger values of $\SGI$ correspond to higher masses.
On the other hand,
with increasing the redshift, larger values of 
$\SGI$ are associated with a given halo mass.
Even though datapoints correspond to different redshifts and cosmologies,
they all form a well defined sequence. This indicates that 
the function $f(\SG)$ is universal to good approximation. 
For a given $\sigma$, outputs at a fixed redshift scatter around the universal
sequence by 10-15 per cent.
A number of fitting formulae have been proposed in the literature to
parameterise this sequence. In Figure \ref{FIG_GAUSSMF},
we compare some of them
(summarised in Table \ref{TAB_MF}) with our datapoints. 
Fractional deviations between models and data are shown in the bottom panel.
Barring the classical Press-Schechter result, all the fitting formulae
describe our data to better than 20 per cent.
The best agreement is found all over the mass range with \citet{WARR} followed 
by \citet{JENK}
which both show deviations from our data at the 10 per cent level. 
The \citet{ST} model also provides an accurate description
of the data for small halo masses
but tends to overestimate the abundance of the most massive objects.
On the other hand, the fit by \citet{REED} tends to underestimate
the high-mass tail of the mass function.
Overall our findings are in good agreement with \citet{HEIT} and \citet{LUK}.

Following \citet{WARR} and \citet{TINK}, we fit the outcome of the 
simulations with the function 
\BE
f(\SG) = \left[ \rm{D} + \rm{B} \left(\frac{1}{\SG}
\right)^{\rm{A}}\right] \exp{\left(-\frac{\rm{C}}{\SG^2}\right)}\;.
\label{EQ_OURFIT}
\EE
The best-fitting parameters have been determined through $\chi^2$ minimisation
using the Markov Chain Monte Carlo method, and read:
\begin{eqnarray}
\label{EQ_BESTGAUFIT}
\nonumber \rm{A} &=& 1.868  \pm   0.019 \\
\rm{B} &=& 0.6853  \pm  0.0035\\
\nonumber \rm{C} &=& 1.2266  \pm  0.0049 \\
\nonumber
\rm{D} &=& 0.2279 \pm  0.0022 \;.
\end{eqnarray}
In terms of the parameterisation given in \citet{WARR} 
and reported in Table \ref{TAB_MF}, this corresponds to
$(\rm{A, a, b,c}) =  (0.6853, 1.868, 0.3324, 1.2266)$.
The fit in equation (\ref{EQ_BESTGAUFIT}) describes our dataset up to 
deviations of a few per cent over the
entire mass and redshift ranges for Run 1.0 and Run 2.0, 
while it shows larger deviations (up to nearly 10 per cent)
towards the high-mass end of Run 3.0, 
(see Figure \ref{FIG_GAUSSMF}). It is important to remember, however, 
that Run 3.0 covers a much smaller volume than the others and 
thus is more severely affected by sample variance.

\subsection{The universal halo mass function from non-Gaussian initial
conditions}
\begin{figure}
  \includegraphics[width=8cm]{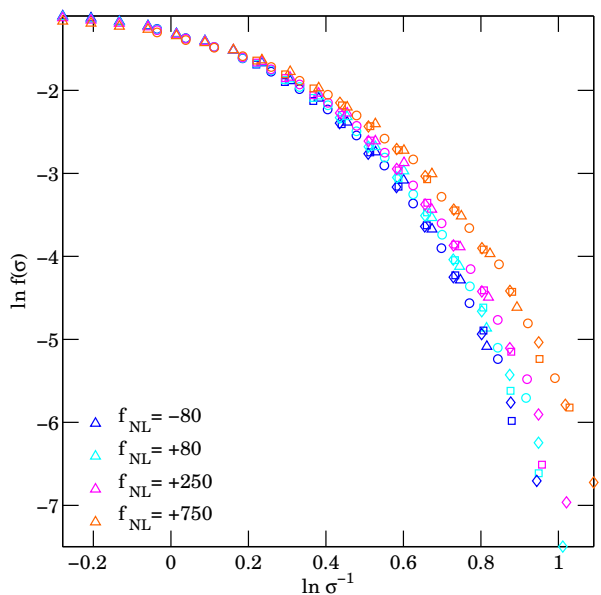}
  \caption{Universality of the mass function arising from non-Gaussian
initial conditions.
Colors refer to simulations with different values of $\FNL$ as indicated by
the labels. Symbols identify the redshift of the simulation output from
which the mass function has been calculated, namely $z=0$ (triangles), 0.5
(circles), 1 (squares), 1.6 (diamonds).}
  \label{FIG_UNIVMF}
\end{figure}

Is the function $f(\SGI)$ universal also in the non-Gaussian case?
This question is addressed 
in Figure \ref{FIG_UNIVMF} where we show the output of our
main
series of simulations at four redshifts ($z=0,0.5,1,1.6$)
to test the scaling of the mass function in terms of $\SGI$.
Only bins containing at least 20 haloes are considered.
Within a certain tolerance, the halo mass functions at different
masses and redshifts all lie on the same curve for a given $\FNL$. 
The scatter of the points at a fixed redshift
around this curve roughly amounts to 10 per cent, and it becomes smaller
towards our largest values of $\FNL$.

We thus
generalise equation (\ref{EQ_MF}) 
to non-Gaussian initial conditions by assuming that
\BE
\frac{dn}{dM}(\FNL,M,z) = f(\FNL,\SG)~ 
\frac{\bar{\rho}_{\rm m}}{M}\frac{d
~\rm{ln}
[\SG^{-1}(M,z)]}{dM}\;,
\label{EQ_NGMF}
\EE
and we provide a fitting formula for $f(\FNL,\SG)$.
Given the similarity to the Gaussian case,
we still adopt the functional form given in equation (\ref{EQ_OURFIT})
but let the parameters $\rm{A,B,C,D}$ vary with $\FNL$.
The best-fitting values have been determined in two steps.
We first used a Markov Chain Monte Carlo method to determine 
$\rm{A,B,C,D}$ at fixed $\FNL$ through $\chi^2$ minimisation.
The results suggest
that the $\FNL$ dependence for each parameter of the mass function 
can be accurately described by polynomials of different orders.
Eventually, we used the data to derive the coefficients of these
polynomials.

\begin{center}
\begin{figure*}
  \includegraphics[width=8cm]{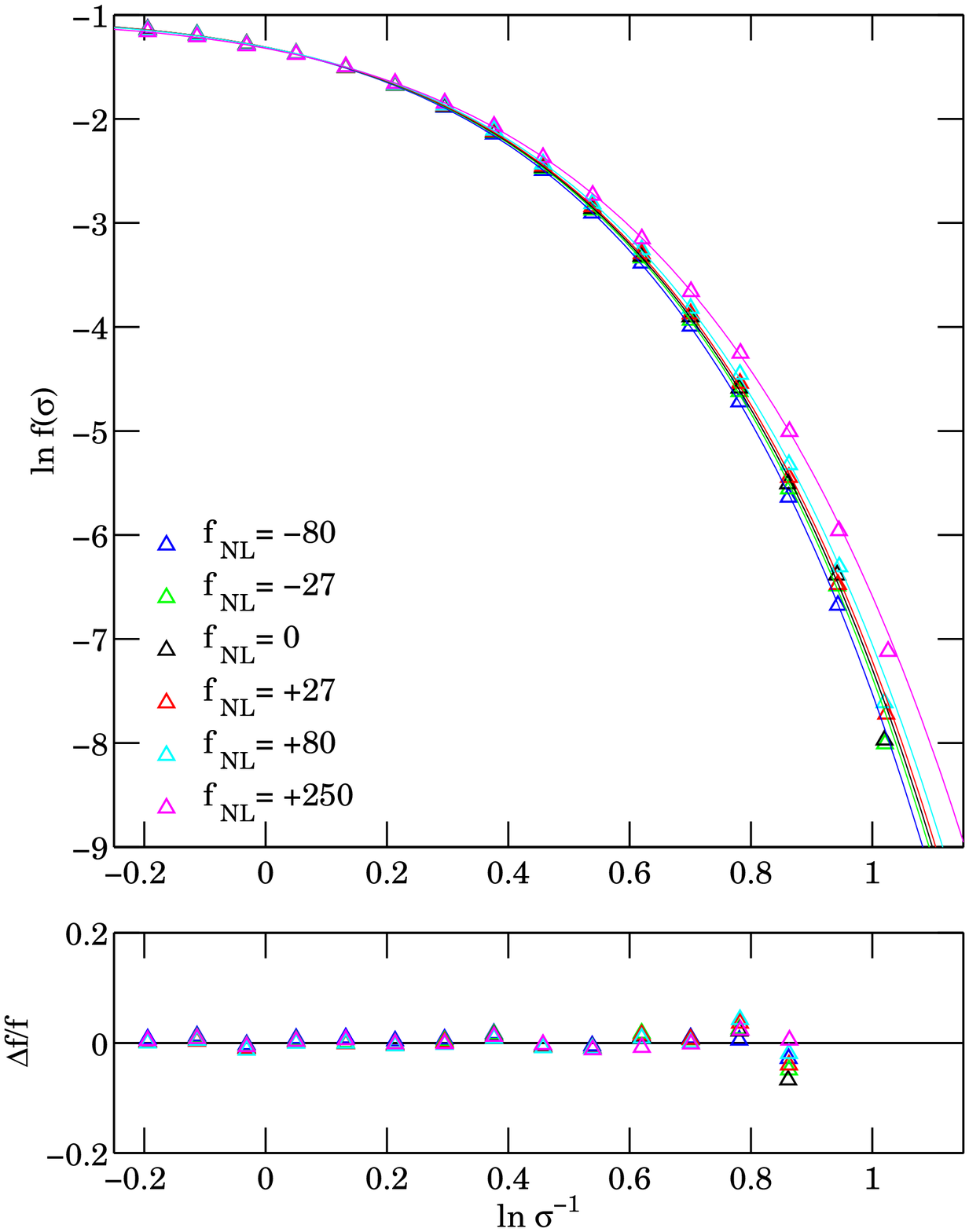}
\includegraphics[width=8cm]{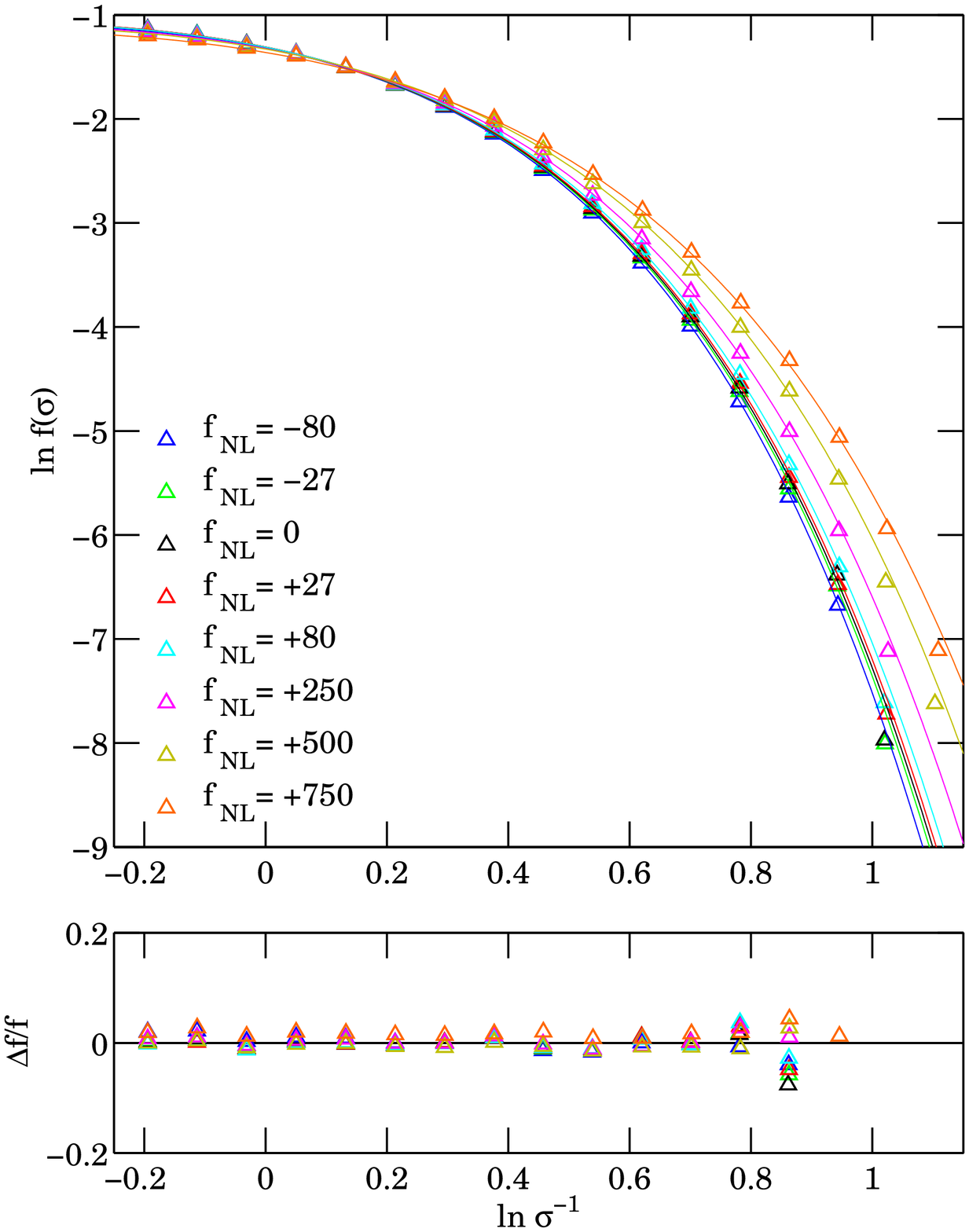}
\caption{Comparison between the halo mass function from our main series
of simulations (triangles) and the corresponding fitting functions (lines). 
Values $-80\leq\FNL\leq 250$ and the fit in equation (\ref{LINEARFIT})
are considered in the left panel. 
All the simulations and the polynomial fit in equations (\ref{EQ_LARGEMFFIT1})
and (\ref{EQ_LARGEMFFIT2})
are shown in the right panel.
The lower panels show
residuals $\frac{\Delta f}{f} = \frac{({\rm data} - {\rm fit})}{\rm data}$ 
for data points with a statistical uncertainty which is smaller
than 5 per cent.}
\label{FIG_MFF}
\end{figure*}
\end{center}
\begin{table}
\begin{center}
\begin{tabular}{|c|cc|}
\hline
Parameter	&  $\rm{p_1}$ & $\rm{p_2}$  \\
\hline
A  &	1.694		&-0.00199\\
B  & 	0.566		&-0.00029\\
C  &	1.151		&-0.00071\\
D  &	0.287		&-0.00030\\
\hline
\end{tabular}
\caption{Best-fitting values for the linear coefficients of the universal
mass-function parameters given in equation (\ref{LINEARFIT}).
The quoted values are truncated at the first digit which is affected by
the statistical errors.
This provides an accurate description of our simulations 
for $-80\leq\FNL\leq 250$.
}
\label{TAB_MFF_SMALL}
\end{center}
\end{table}

\begin{table}
\begin{center}
\begin{tabular}{|c|ccccc|}
\hline
Parameter	&  $\rm{p_1}$ & $\rm{p_2}$&$\rm{p_3}$&$\rm{p_4}$&$\rm{p_5}$ \\
& &$(10^{-3})$&$(10^{-7})$&$(10^{-9})$&$(10^{-12})$\\
\hline
A	&1.708   &-2.07   & 3.1		&0	&0	\\
B	&0.560   &-0.46   & +12.46	&-2.36	&+2.65	\\
C	&1.150   &-0.76   & +2.7	&0	&0	\\
D	&0.293   &-0.16   &-14.07	&+3.88   &-3.94	\\
	\hline
\end{tabular}
\caption{As in Table \ref{TAB_MFF_SMALL} but for the fitting formula
in equations (\ref{EQ_LARGEMFFIT1}) and (\ref{EQ_LARGEMFFIT2}). This accurately
describes the mass
function in all our non-Gaussian simulations ($-80\leq\FNL\leq750$).}.
\label{TAB_MFF_LARGE}
\end{center}
\end{table}

The degree of complexity required to fit the simulation data grows
considerably with increasing $\FNL$.
For $-80\leq\FNL\leq 250$ (a range that fully encloses the values currently
allowed by CMB studies), the mass-function parameters in equation
(\ref{EQ_OURFIT}) are well approximated by the linear relation
\BE
\rm{P}(\FNL) = \rm{p}_1 + \rm{p}_2 \cdot \FNL,\ \ \  ~for~ \rm{P= A,B,C,D}.
\label{LINEARFIT}
\EE 
Table \ref{TAB_MFF_SMALL} lists the corresponding best-fitting parameters.
The quality of this fitting formula is assessed in 
the left panel of Figure \ref{FIG_MFF}, where the mass function for the
simulations with $\FNL=-80, -27,0,+27,+80,+250$ is compared with
the corresponding fit. Residuals (shown in the bottom panel) 
are smaller than 5 per cent all over the range $ -0.2 < \rm{ln}\,\SGI < 0.8$
corresponding to the mass interval
$2\times10^{13}<M<2\times10^{15} \HI \MSUN$ at $z=0$.\\
On the other hand, equation (\ref{LINEARFIT})
is not suitable to account for values of $\FNL$ substantially larger than 250.
To obtain an accurate fit of the universal halo mass function over the range 
$-80\leq \FNL\leq 750$ we had to consider 
polynomials up to $4^{\rm th}$ order in $\FNL$:
\BE
\rm{P}(\FNL) = \rm{p}_1 + \rm{p}_2 \cdot \FNL + \rm{p}_3 \cdot \FNL^2, \ \ \
~for~
\rm{P= A,C}
\label{EQ_LARGEMFFIT1}
\EE 
and
\begin{eqnarray}
\label{EQ_LARGEMFFIT2}
\rm{P}(\FNL) = \rm{p}_1 + \rm{p}_2 \cdot \FNL + \rm{p}_3 \cdot \FNL^ 2+ \rm{p}_4
\cdot \FNL^3 \!\!\!\!\!\!&+&\!\!\!\!\!\! \rm{p}_5 \cdot \FNL^4 , \\
 &{\rm for}& \rm{P= B,D}. \nonumber
\end{eqnarray}
The best-fitting values of the parameters above are listed in Table
\ref{TAB_MFF_LARGE} while the corresponding functions are compared with
the simulation data in the right panel of 
Figure \ref{FIG_MFF}. Also in this case residuals are smaller than
5 per cent for $\rm{ln}\,\SGI < 0.8$.\\

The universality of the fitting formula in equation (\ref{EQ_NGMF})
has been further tested against our non-Gaussian 
simulation of the WMAP3 cosmology, 
Run2.750, which has not been used
to determine the best-fitting parameters. This blind check shows that,
in the range $ -0.27< \rm{ln}\,\SGI < 0.94$ (roughly corresponding to
$1.6 \times 10^{13} < M < 2.2 \times 10^{15}\,\HI \MSUN $ at $z=0$), 
the provided fit
reproduces the mass function with an accuracy of 5 per cent.

We warn the readers against extending our fitting formulae beyond their
range of validity, in particular at low halo masses.
The simulations of our main series resolve $10^{13}\, \HI M_\odot$ haloes
with 100 particles. For $\FNL \neq 0$, 
our analytical formulae for the mass function 
have been derived using only haloes that are more massive than this limit.
Moreover, since the high-mass tail of the mass function is enhanced (suppressed)
for positive (negative) values of $\FNL$ with respect to the Gaussian case,
mass conservation requires that the opposite effect is seen at lower masses.
We have directly tested the goodness of our fit towards the smaller masses 
using Run3.250 (which has a boxsize
8 times smaller than for the simulations in the main series but the same
number of particles) and indeed found
that the fitting formulae in equations (\ref{EQ_LARGEMFFIT1}) 
and (\ref{EQ_LARGEMFFIT2})
systematically overestimate the abundance of small mass haloes by 10-30 
per cent. 
We will address the low-mass tail of the mass function for
$\FNL\neq0$ in future work.

On the other hand, for Gaussian initial conditions, we combined simulations
with different box sizes to derive the fitting function in equations
(\ref{EQ_OURFIT}) and
(\ref{EQ_BESTGAUFIT}). This allowed us to extend the validity 
of our fit to the much wider mass range
$2.4\times 10^{10}<M<5\times 10^{15}\,\HI M_\odot$. \\
Our fitting formulae give three different approximations for
the universal mass function in the Gaussian case. In general, the fit given
in equations (\ref{EQ_OURFIT}) and
(\ref{EQ_BESTGAUFIT}) has to be preferred 
as it has been obtained from
a richer dataset spanning a much wider range of halo masses.
However for masses
above $10^{13} \HI M_\odot$ at $z=0$, the fit in equations
(\ref{EQ_OURFIT}) and (\ref{LINEARFIT}) and Table \ref{TAB_MFF_SMALL} provides
the most accurate representation of our
data. In any case, the different
fitting functions never deviate by more than 3-4 per cent.
Also note that our two fitting functions for the non-Gaussian simulations 
agree by better than 1 per cent for $-27\leq\FNL\leq80$ and by 
a few per cent for $\FNL=-80$ and $\FNL=+250$.
\begin{center}
\begin{figure*}
\includegraphics[width=8cm]{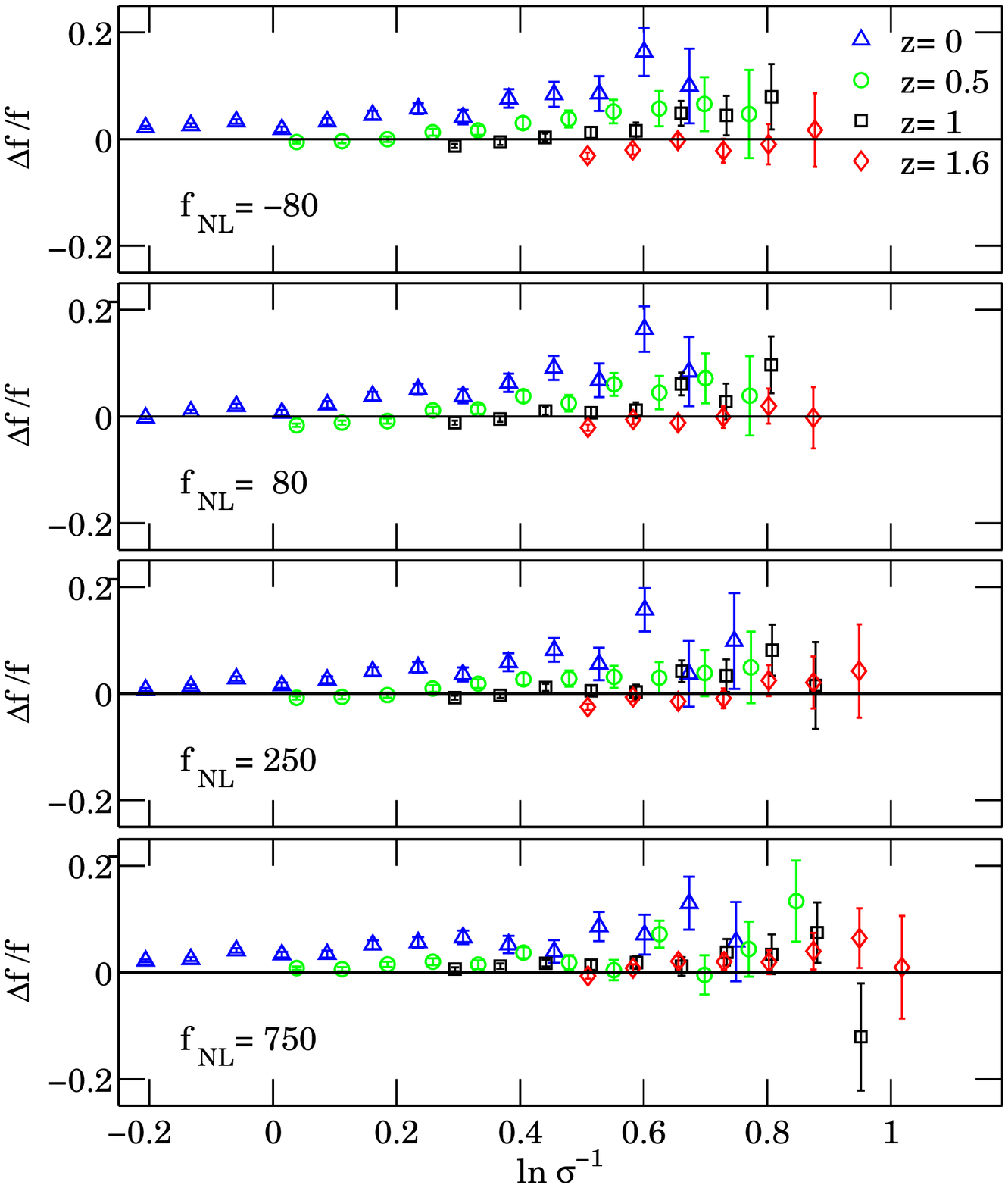}
\includegraphics[width=8cm]{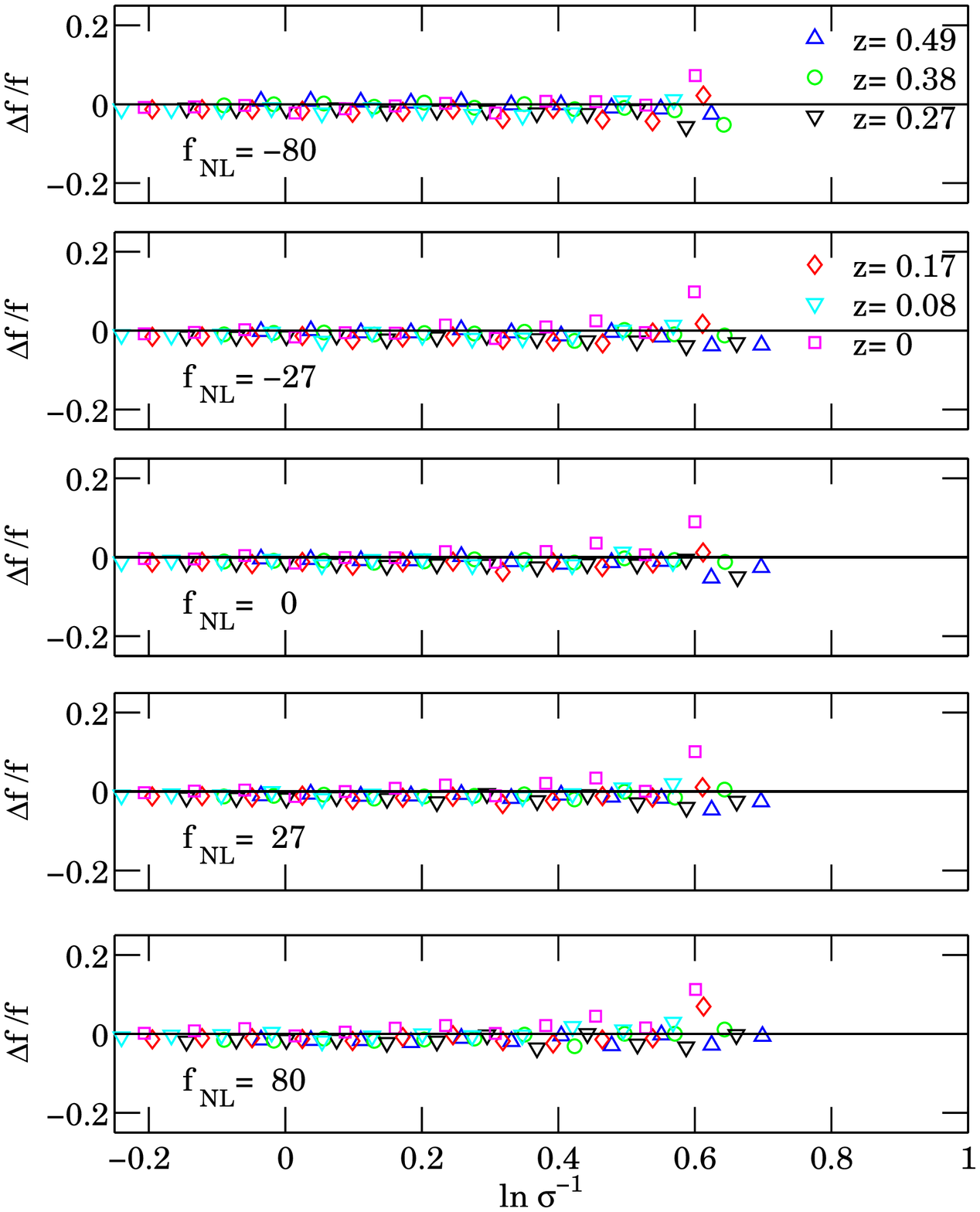}
\caption{Left panel: Mass function residuals of Run1.-80, Run1.80, Run1.250, 
Run1.750 with respect to the universal fit given in equations
(\ref{EQ_LARGEMFFIT1})
and (\ref{EQ_LARGEMFFIT2}) 
at redshift $z = 0,~ 0.5,~ 1, ~1.61$ 
(indicated by the symbols and colors). 
Only data points with a statistical error smaller than 10 per cent are shown.
Right panel: As in the left panel but for the non-universal fit given in
equation (\ref{EQ_ZANDFNLFIT}). 
In this case, only data points with an accuracy better than 5 per cent are 
shown.}
\label{FIG_UNIVERSALITYLIMITS}
\end{figure*}
\end{center}
\subsection{The limit of universality: redshift dependence}
Regardless of the value of $\FNL$,
we have found that the halo mass function is universal, when written in
terms of $\SGI$, with an accuracy of roughly 10 per cent. 
If one is interested in giving
analytical approximations for the halo mass function which are more accurate
than the universal fit,
it is necessary to introduce redshift-dependent corrections
(see also \citealt{TINK} for the Gaussian case).
In the left panel of Figure \ref{FIG_UNIVERSALITYLIMITS}, we show
how well the universal fit 
(whose parameters are listed in Table \ref{TAB_MFF_LARGE}) describes 
the simulation outputs at $z = 0,~0.5,~1,~1.61$.
At $z=0$ and
for masses $M \geq 4-5 \cdot 10^{14} \MSUN$ 
the fitting formula deviates for the data by more than
10 per cent. The smaller the redshift, the worse is the agreement between the 
data points and the universal fit. 
The bigger the $\FNL$, the less critical is the comparison.\\
In this Section we provide a non-universal fit which is very accurate at
low redshift. In particular, we write:
\BE
\frac{dn}{dM}(\FNL,M,z) = f(\FNL,\SG_0, z)~ 
\frac{\bar{\rho}_{\rm m}}{M}\frac{d
~\rm{ln}
[\SG^{-1}_0(M)]}{dM}\;,
\label{EQ_NGMFZ}
\EE
where $\SG_0 = \SG(z=0) = \SG(z)/ D_{+}(z)$ is the rms deviation of the
linear density field at $z=0$.
We approximate $f$ with
the functional form given in equation (\ref{EQ_OURFIT})
but now let the parameters $\rm{A,B,C,D}$ vary with both $\FNL$ and $z$.
Markov Chain Monte Carlo fitting
suggests that each parameter $\rm{A,B,C,D}$ of the mass function can be 
accurately described as follows:
\BE
\rm{P}(z, \FNL) = \rm{p}_1 ~ [1+\rm{p}_2 \cdot z+\rm{p}_3 \cdot
  z^2]~[1+\rm{p}_4 \cdot \FNL]\;.
\label{EQ_ZANDFNLFIT}
\EE
\begin{table}
\begin{center}
\begin{tabular}{|c|cccc|}
\hline
Parameter	&  $\rm{p_1}$ & $\rm{p_2}$&$\rm{p_3}$&$\rm{p_4}$ \\
& &$(10^{-1})$&$(10^{-1})$&$(10^{-4})$\\
\hline
A & 1.82  &  2.85  &  4.53  & -5.92 \\
B & 0.578 &  5.30  &  7.53  & -7.73 \\
C & 1.15  &  9.52  &  9.08  & -4.42 \\
D & 0.294 &  4.92  &  4.67  & -3.80 \\
\hline
\end{tabular}
\caption{Best fitting values for the mass-function parameters given in equation
(\ref{EQ_ZANDFNLFIT}). 
This provides an accurate description of the data for $0 \le z \le 0.5$ and $-80
\le \FNL \le 80$.
The universal function of equations (\ref{EQ_NGMF}), (\ref{EQ_OURFIT}),
(\ref{EQ_LARGEMFFIT1}). and (\ref{EQ_LARGEMFFIT2}) should be used otherwise.}
\label{TAB_ZANDFNLFIT}
\end{center}
\end{table}
The best fitting parameters 
for $-80\leq\FNL\leq 80$ and $0\leq z \leq 0.5$ 
are listed in Table \ref{TAB_ZANDFNLFIT}, 
while the quality of the fitting formula is assessed in the right panel of
 Figure \ref{FIG_UNIVERSALITYLIMITS}. Residuals are smaller 
than 5 per cent all over the mass range, indicating that for 
$-80\leq\FNL\leq 80$ and $0\leq z \leq 0.5$ the fit of equations 
\ref{EQ_NGMFZ}, \ref{EQ_OURFIT}, and \ref{EQ_ZANDFNLFIT} has to be preferred 
to the universal fit given in the previous Section. On the other hand, for
higher values 
of $|\FNL|$ and for higher redshifts, the universal fit 
gives a better and more economic (in terms of parameters) description of the 
data.

\begin{center}
\begin{figure*}
\includegraphics[width=8cm]{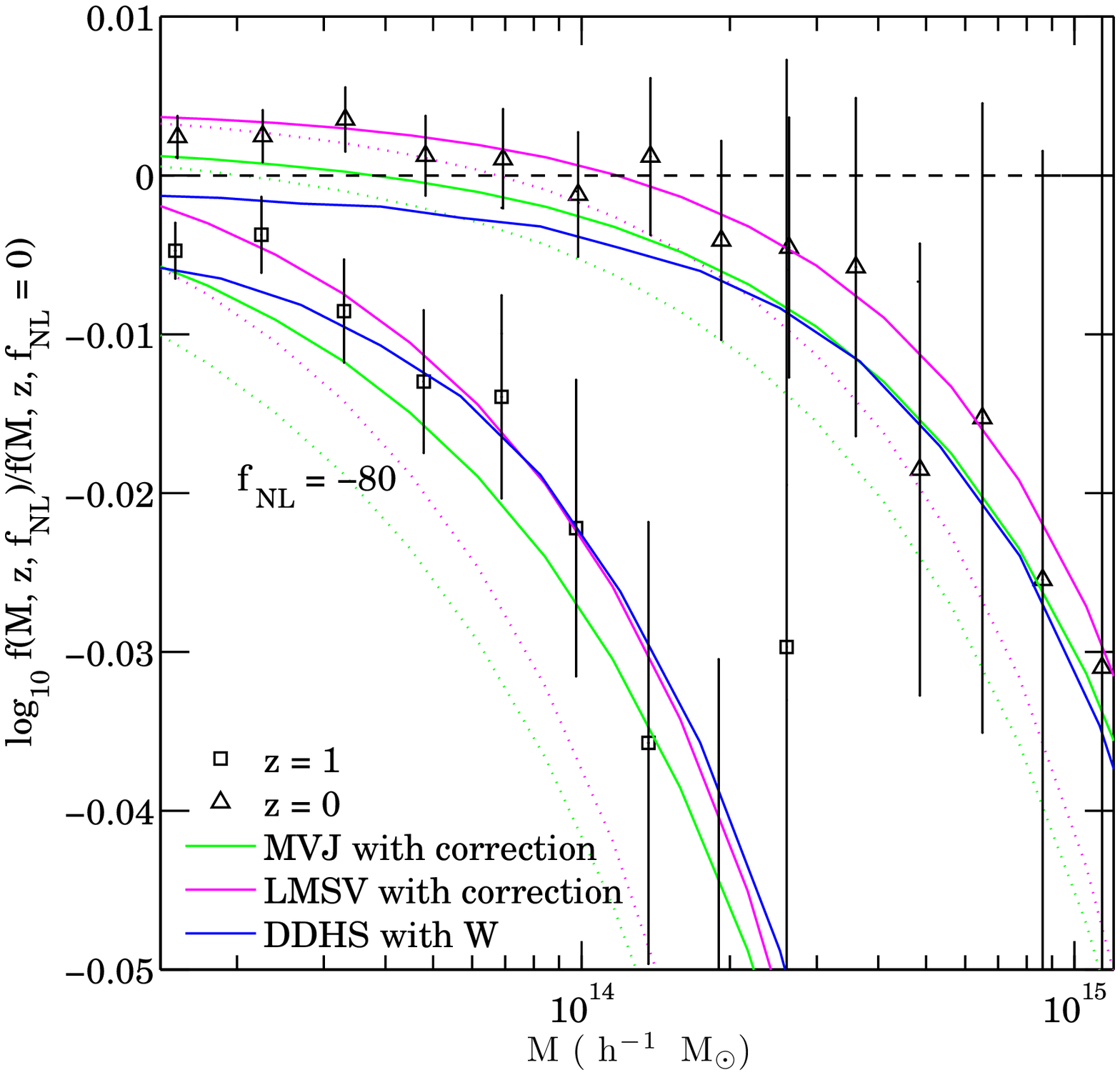}
\includegraphics[width=8cm]{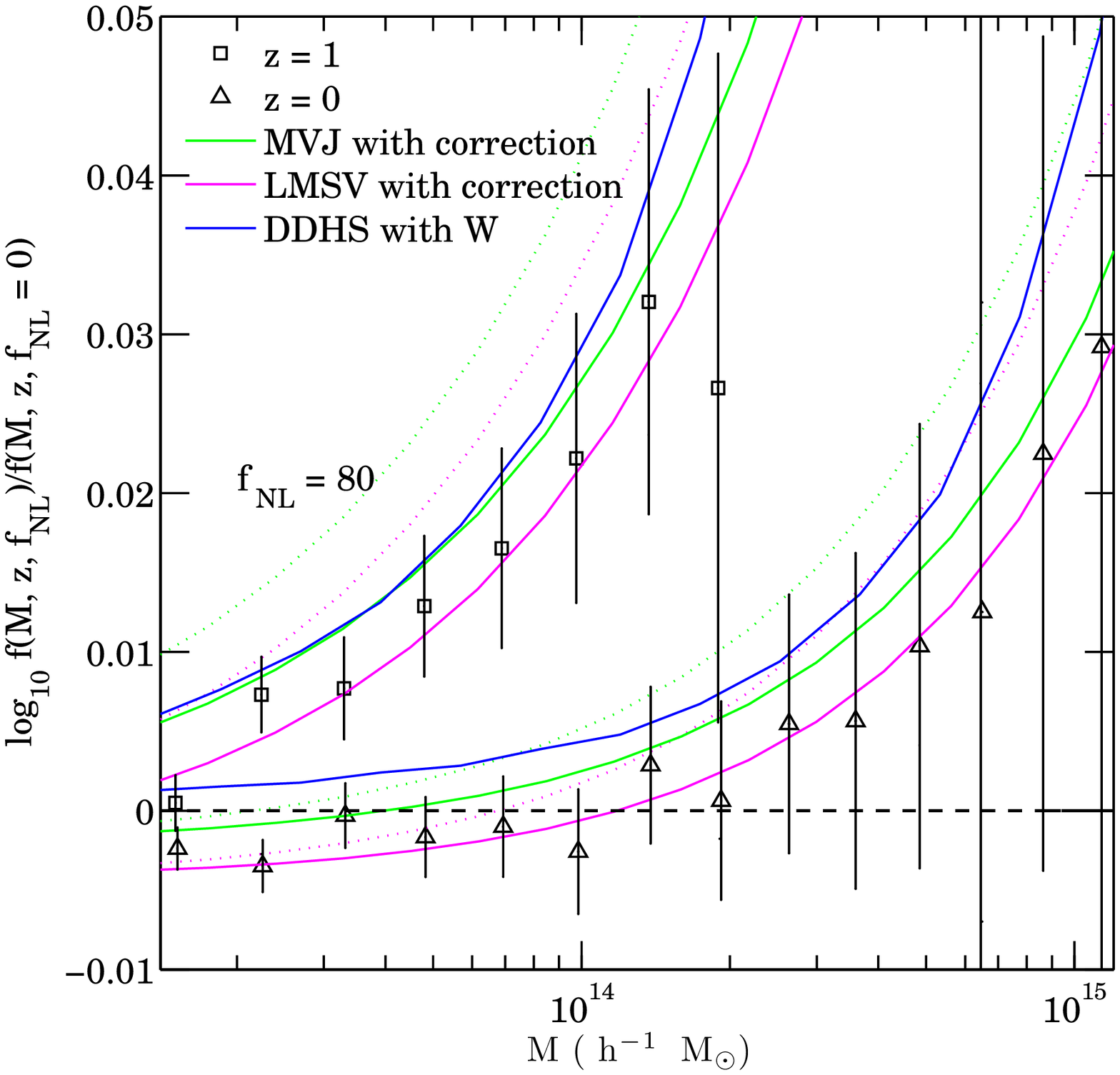}
\includegraphics[width=8cm]{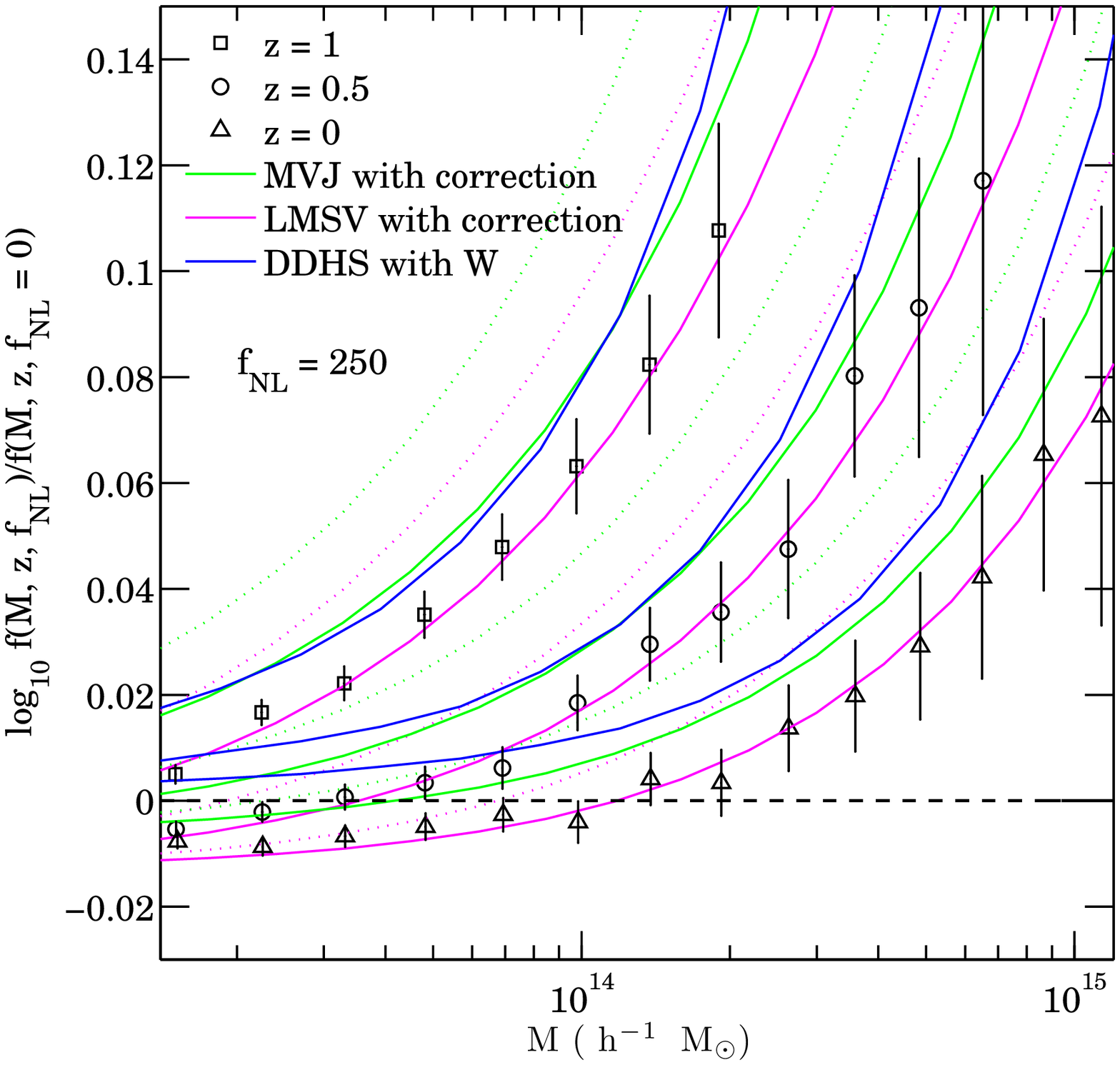}
\includegraphics[width=8cm]{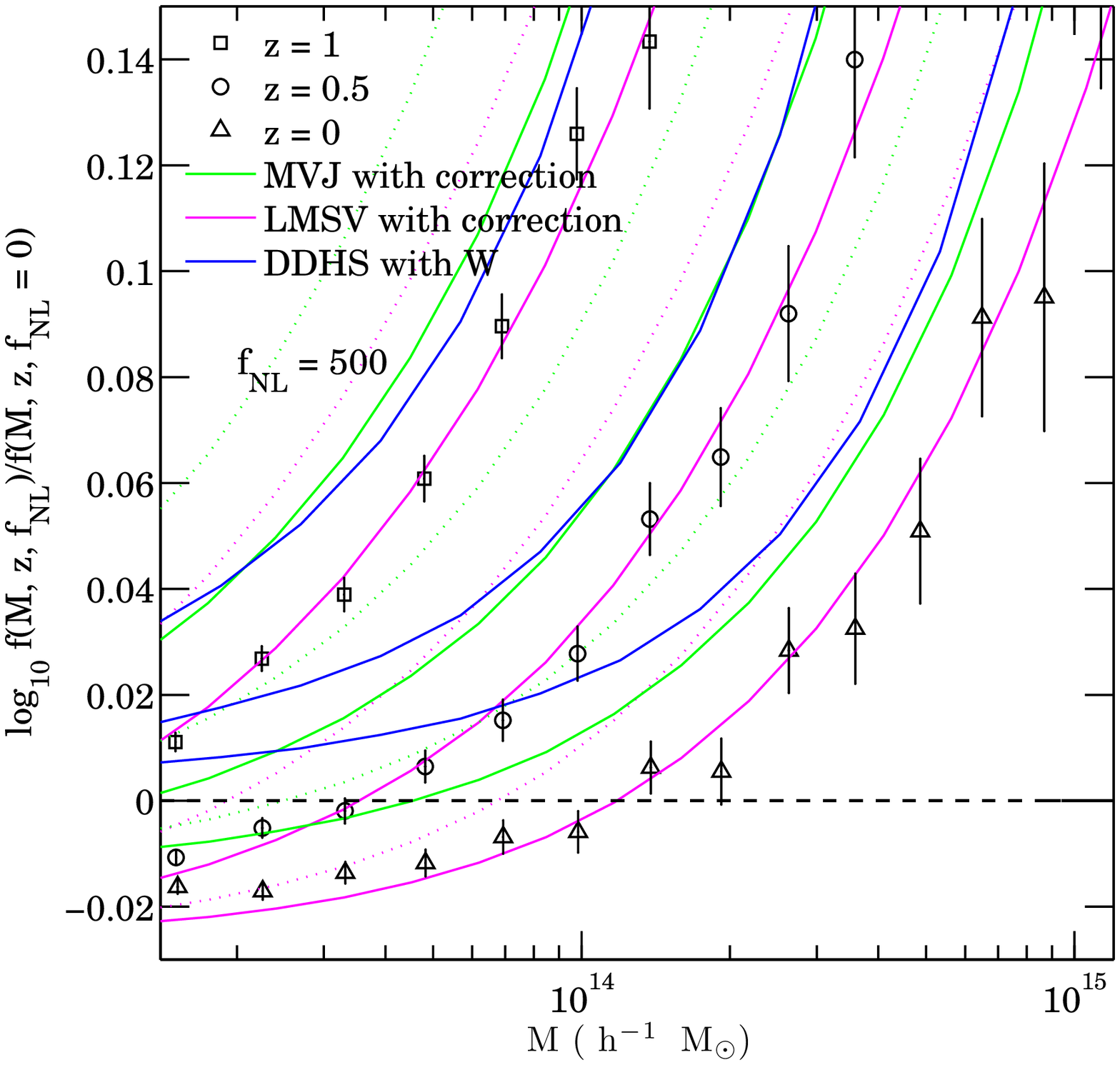}
\caption{Comparison between the halo mass functions from our
simulations and from the models by \citet{MVJ}, by \citet{LOVERDE}, and the fit
by \citet{DALAL} for different values of $\FNL$ (different panels) and for
$z=0,~
0.5,~ 1$ (triangles, circles, squares, respectively). The quantity which
is plotted is the ratio $f(z, \FNL)/f(z,\FNL=0,)$. 
The dotted lines indicate the models of
\citealt{MVJ} (green) and \citealt{LOVERDE} (magenta), as
they appear in equations (B.6) and (4.19) of \citet{LOVERDE}, respectively.
The corresponding solid lines indicate the same models with a reduced threshold for halo collapse:
$\DELTAC \simeq 1.5$.
The blue solid lines are obtained by convolving the $\FNL$-dependent
kernel given in \citet{DALAL} with the mass-function fit for
the Gaussian case by \citet{WARR}.}
\label{FIG_MFCOMP_RATIO}
\end{figure*}
\end{center}
\begin{center}
\begin{figure*}
\includegraphics[width=8cm]{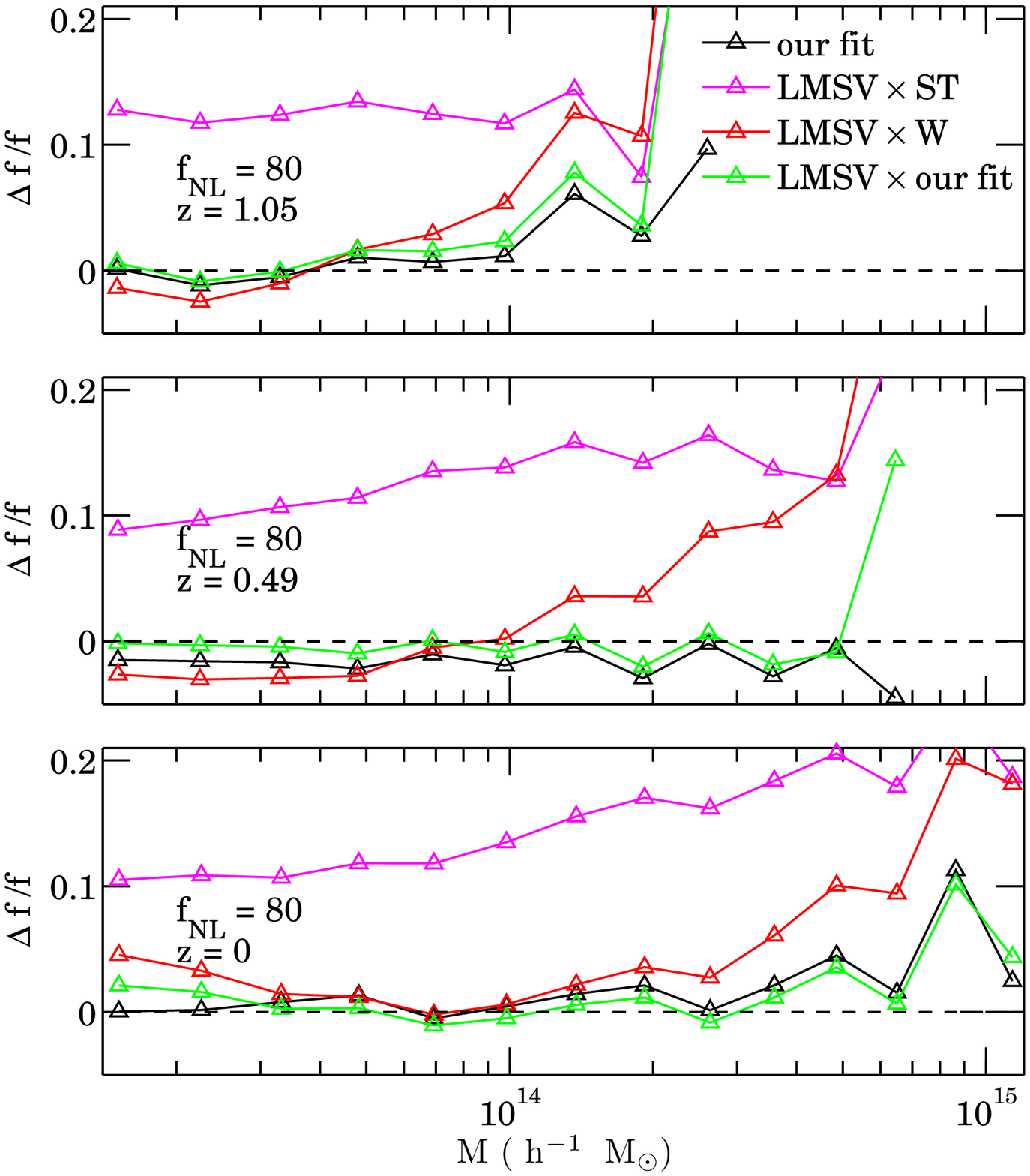}
\includegraphics[width=8cm]{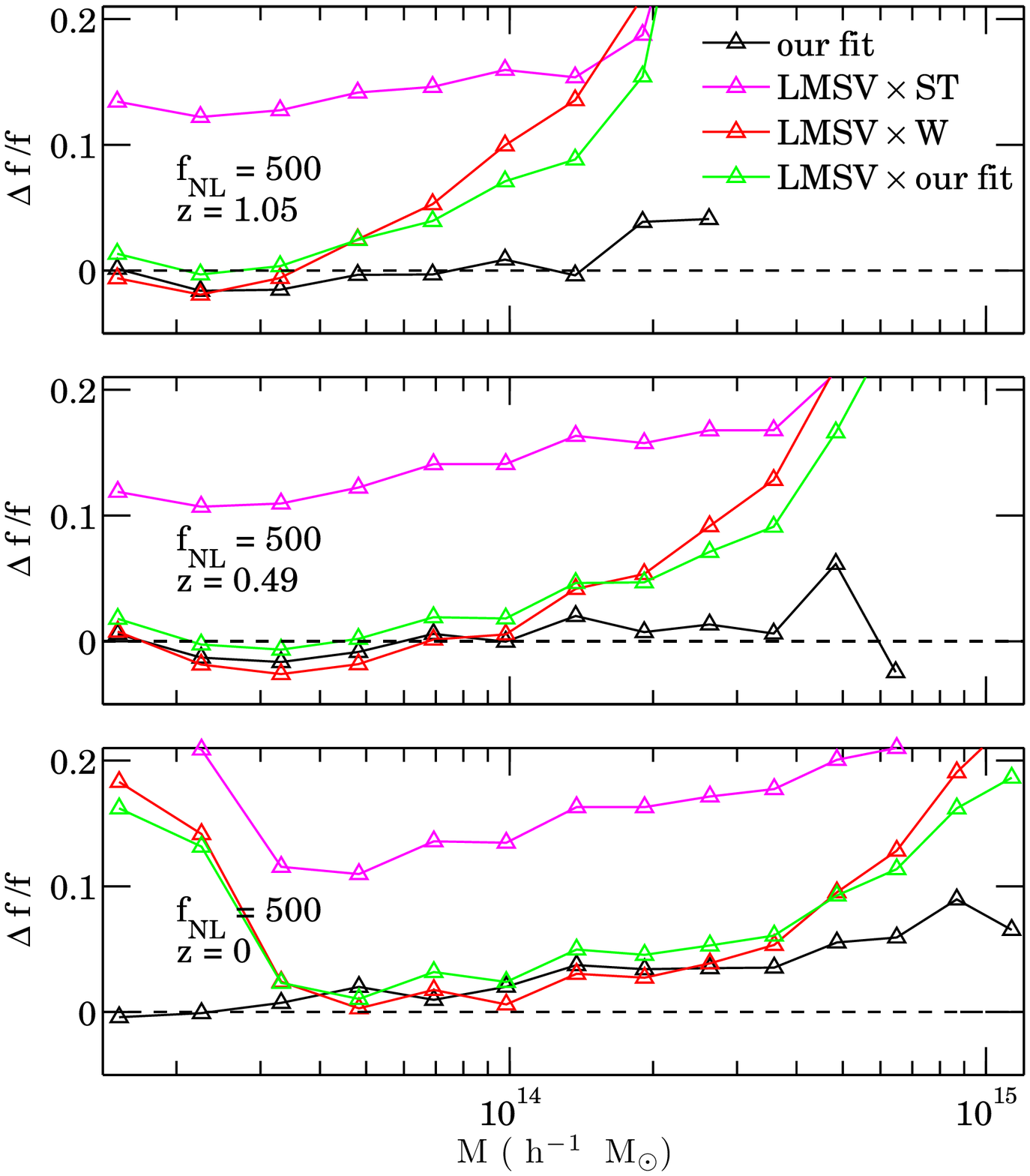}
\caption{ Residuals 
between the simulated mass functions and various model prescriptions,
for Run1.80 (left-hand panels) and Run1.500 (right-hand panel) at $z=0,~0.5,~ 1
$. The models have been obtained multiplying the 
formula for $f(z, \FNL)/f(z,\FNL=0,)$ by \citet{LOVERDE} (with the reduced
$\delta_{\rm c}$ as the solid lines in Figure \ref{FIG_MFCOMP_RATIO}) with 
different Gaussian mass functions: 
\citealt{ST} (magenta), \citealt{WARR} (red), and
our Gaussian fit (green). The black lines show residuals with respect to
our fitting formula given in Sections 3.2 and 3.3.
Only data with errors smaller than 10 per cent are
shown.}
\label{FIG_MFCOMP_RES}
\end{figure*}
\end{center}
\subsection{Comparison with theoretical models}
The halo mass function arising from mildly non-Gaussian
initial conditions can be modelled by generalizing the Press-Schechter 
formalism.
Using the saddle-point approximation to evaluate the probability for the
linear density field to be above a given threshold value,
\citet{MVJ} have derived a model for $dn/dM$.
More recently, \citet{LOVERDE} presented another expression for the mass
function by using 
the Edgeworth asymptotic expansion for the probability density function
of the linear density field.
In both cases, only leading-order corrections in $\FNL$ have been
accounted for. 
In absolute terms, these models are not expected to be accurate as 
they should suffer from the same shortcomings as the Press-Schechter model
in the Gaussian case.
However, they can be used to compute the fractional non-Gaussian correction 
$f(M,z, \FNL)/f(M,z,\FNL=0)$ \citep{VWHK, LOVERDE}. 
In Figure \ref{FIG_MFCOMP_RATIO}, we use this quantity
to test the models against our simulations. 
Datapoints with errorbars show
the N-body output at $z=0, 0.5$ and 1, while
the dotted lines indicate the models of \citet{MVJ} and \citet{LOVERDE} 
as indicated in equations (B.6)\footnote{This fixes a typo in equation
(68) of \citet{MVJ}} and (4.19) of \citet{LOVERDE},
respectively.
It is evident that the models overestimate the non-Gaussian correction.
Following the indications in \citet{CVM}, 
we also show a modified version of the models which is obtained by lowering
the critical threshold for halo collapse as 
$\DELTAC \simeq 1.5$
(solid lines in Figure \ref{FIG_MFCOMP_RATIO}).
Such a correction vastly improves the agreement with the simulations. \\
\citet{DALAL} proposed to fit the halo mass function in terms
of the convolution between $dn/dM(\FNL=0,M,z)$ and a Gaussian kernel in
$M$ with a $\FNL$-dependent mean and variance. Figure \ref{FIG_MFCOMP_RATIO} shows that their
fit tends to overestimate the non-Gaussian corrections especially for
large, positive values of $\FNL$ and masses $M<10^{14} h^{-1} M_\odot$.
On the other hand, for $|\FNL|<100$ it has a similar accuracy as the
formulae derived from the Press-Schechter formalism corrected with the
reduced threshold.\\

The good agreement between the fractional non-Gaussian corrections derived from
the modified PS models and from the simulations is not enough to derive 
$\FNL$ from future observations of galaxy clusters.  
In fact the ratio $f(z, \FNL)/f(z,\FNL=0)$ is not an observable:
the only quantity that we can hope to 
compare with observations is the mass function. 
In order to make predictions for $dn/dM$,
the models for the fractional non-Gaussian correction need 
to be multiplied by a Gaussian mass function. This step might introduce
relatively large systematic errors (see Figure 1) which could degrade
any measurement of $\FNL$ based on the cluster mass function.
We address this issue in Figure \ref{FIG_MFCOMP_RES}
where we plot the fractional deviation of some model predictions for
the function $f$ with respect to the simulation output (results are
very similar for different values of $\FNL$).
We consider the model by \citet{LOVERDE} corrected with the factor $\mathcal{N}$
and multiplied by three different Gaussian models: \citet{ST}, \citet{WARR}, 
and our fit with $\FNL=0$.
Note that some of the final outcomes systematically differ by 10-20 per cent 
over the entire mass range covered by the simulations. This clearly shows 
that a careful measurement of the Gaussian mass function is necessary to avoid
a biased estimation of the non-linearity parameter.
Note that, for $|\FNL|<100$, the models by \citet{MVJ} and \citet{LOVERDE}
(both with the reduced collapse threshold) combined with our Gaussian fit  
are in rather good agreement with the numerical mass functions (similar
results are obtained using the Gaussian fit by \citet{WARR} for masses below a
few
$\times 10^{14} h^{-1} M_\odot$).
Perhaps not surprisingly, no model describes the simulation data for all the
values of $\FNL$ as well as our
fitting formulae for the non-Gaussian mass function given in Sections 3.2 and
3.3.
\subsection{Summary of accuracy and range of validity of the mass function fits}
In order to facilitate the use of our fitting formulae for
the halo mass function we summarize here their accuracy and range
of validity.
\begin{itemize}
 \item For $-80\leq\FNL\leq 80$ and $0\leq z \leq 0.5$ the best
description (with 5 per cent accuracy) of our numerical data is given by 
equations (\ref{EQ_NGMFZ}), (\ref{EQ_OURFIT}) and
 (\ref{EQ_ZANDFNLFIT});
 \item For larger values of $\FNL$ and $z$ (but with $\FNL \le 750$ and $z \le
1.6$) or whenever an accuracy of 10 per cent is enough, 
the universal fits of Section 3.2 should be used:
\begin{itemize}
\item universal fit for $-80\leq\FNL\leq 250$: equations (\ref{EQ_OURFIT}),
(\ref{LINEARFIT}) and Table \ref{TAB_MFF_SMALL};
\item universal fit for $-80\leq\FNL\leq 750$: equations (\ref{EQ_OURFIT}),
(\ref{EQ_LARGEMFFIT1}), (\ref{EQ_LARGEMFFIT2}) and Table \ref{TAB_MFF_LARGE}.
\end{itemize}
\end{itemize}
\section{Matter Power Spectrum}
\begin{figure}
\includegraphics[width=8cm]{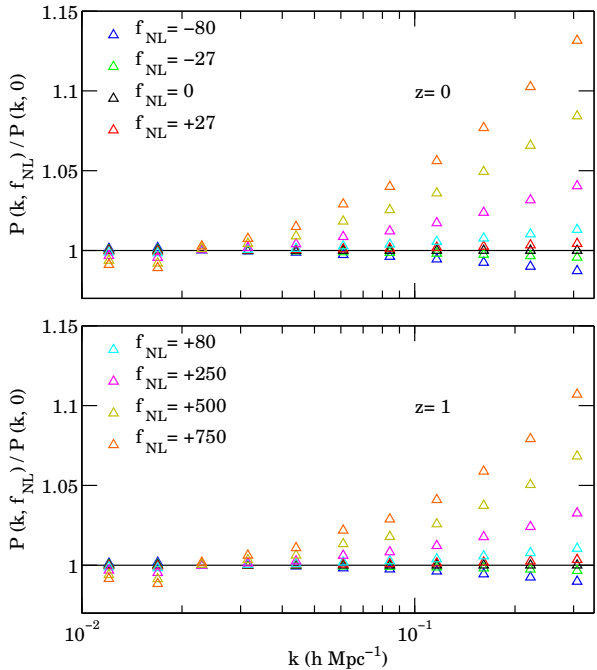}
\caption{Ratio between the matter power spectra of non-Gaussian and 
Gaussian simulations at redshift $z=1$ (bottom) and $z=0$ (top).
Data are extracted from the N-body simulations in our main series
where identical random phases have been used to generate $\phi$ for all
values of $\FNL$. 
}
\label{FIG_POWERSPECTRUM}
\end{figure}

In this section we study how non-Gaussian initial conditions
influence the power spectrum of the mass density field. 
At tree level, the power spectrum does not depend on $\FNL$ in
Eulerian perturbation theory. However, one-loop corrections 
make the power spectrum $\FNL$-dependent.
Qualitatively, theoretical expectations are that
positive (negative) values of $\FNL$
tend to enhance (suppress) the amplitude of the power spectrum on non-linear
scales. 
In Figure
\ref{FIG_POWERSPECTRUM} we plot the ratio of power spectra 
$P(k,\FNL)/P(k,\FNL=0)$ extracted from the simulations
of our main series at redshifts $z=0$ and 1.
The matter power spectrum of non-Gaussian models appears to deviate 
already by a few per cent at $k=0.1\, h$ Mpc$^{-1}$. As expected, deviations
become more severe with increasing the wavenumber $k$.
Our results are in agreement with the perturbative
calculations 
by \cite{TKM}. 
We note, however, that \cite{GROSSI2} found
smaller deviations between the non-Gaussian and Gaussian power spectra 
at larger values of $k$ and $\FNL$.\\

Our results have two important practical implications.
First, the widespread habit of using the Gaussian matter power spectrum
to determine non-Gaussian bias parameters leads to scale-dependent
systematic errors that might become severe when high-precision is required.
Second, primordial non-Gaussianity modifies the power-spectrum
on the scales where baryonic oscillations (BAOs) 
are present. Reversing the argument, two-point statistics
could be also used to constrain the value of $\FNL$. 
Note however, that all probes based on galaxy clustering
will suffer from uncertainties in the bias parameter (and its scale dependence)
that might hinder a measure of $\FNL$ based on the study of BAOs.
On the other hand, weak lensing studies will directly measure the matter
power spectrum.
The target of many future wide-field missions is to provide estimates 
at the per cent level. For parameter estimation,
a comparable accuracy will be required on model spectra within a wide range
of wavenumbers centred around $k\sim 1\ h$ Mpc$^{-1}$ \citep{HT}. 
Therefore, even values of $\FNL$ within the current CMB constraints
could imprint detectable effects in the matter power spectrum
at the scales of interest.
The key question is whether one can discern the effect
of $\FNL$ and, consequently, how much primordial non-Gaussianity will affect the
estimate of the other cosmological parameters.
We will get back to this in future work.

\section[]{Halo Clustering}

The clustering of dark-matter
haloes is biased relative to that of the underlying mass distribution
by an amount which depends on halo mass, redshift, and the
scale at which the clustering is considered \citep[see e.g.][]{MOWHITE,CMP,SSS}.
For Gaussian initial conditions,
this has been widely tested against numerical simulations
\citep[e.g.][]{SMT,SW,TWZZ}.

\begin{figure}
\includegraphics[width=8cm]{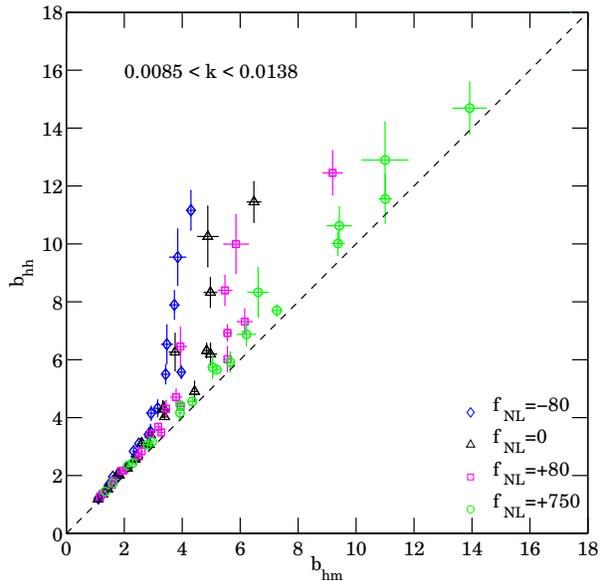}
\caption{The halo bias from the halo-halo power spectrum (with no discreteness
corrections) is plotted against the halo bias 
from the halo-matter cross spectrum.
Whenever the density of haloes is high enough, the two estimates are very close
showing that little stochasticity between mass and halo overdensities
is present on the scales of interest (indicated in $h$ Mpc$^{-1}$ in the label).
The excess in $b_{\rm hh}$ for rare, massive haloes is likely due to shot noise.
Note that large positive values of $\FNL$ correspond to more massive haloes
and thus allow more accurate measures of high bias parameters.}
\label{BHHBHM}
\end{figure}

In general, the halo bias can be quantified 
using either the power spectrum of the halo density field, $P_{\rm hh}$,
or the cross-spectrum between the halo and the
underlying matter density field, $P_{\rm hm}$. 
In the two cases the bias reads
\BE
b_{\rm hh}(k,M,z) = \sqrt{\frac{P_{\rm hh}(k,M,z)}{P(k,z)}}\;,
\label{BHHDEF}
\EE
or 
\BE
b_{\rm hm}(k,M,z) = \frac{P_{\rm hm}(k,M,z)}{P(k,z)}\;,
\label{BHMDEF}
\EE
where $P(k,z)$ is the matter power spectrum. 
If the bias due to halo formation is local and deterministic 
then $b_{\rm hh}=b_{\rm hm}$ apart from measurement errors.
However, in the presence of a stochastic component 
that does not correlate with the density field
$b_{\rm hh}\geq b_{\rm hm}$.
In practice, however, the measurement of all power spectra
is affected to some level 
by shot noise due to the discrete nature of dark-matter
haloes and N-body particles. 
If the distribution of the tracers can be approximated as the Poisson sampling
of an ideal density field, then the measured power spectrum
corresponds to that of the underlying field 
plus the mean volume per particle \citep{PEEBLES}.
Discreteness effects are thus expected to be negligible for $P$ and $P_{\rm hm}$
due to the large number density of particles in the simulations.
On the other hand, massive haloes are rare and, being extended objects,
cannot be modelled as the Poisson sampling of a continuous distribution
(\citealt{MOWHITE}, \citealt{MP}, Porciani in preparation).
It is not clear then how to correct for the discreteness effect in their power
spectrum \citep{SSS}.
For these reasons we use $b_{\rm hm}$ in our analysis
and we adopt $b_{\rm hh}$ (without performing any
discreteness correction) only to verify the results
(see Figure \ref{BHHBHM}).

\subsection{Halo bias from Gaussian initial conditions}
It is well known that the halo bias factor from Gaussian
initial conditions is approximately scale-independent for small 
values of the wavenumber $k$.
We will refer to this asymptotic value on large scales as
the ``linear bias'' and denote it by $b_0$.
Similarly to the halo mass function,
when expressed in terms of $\SGI$, the linear bias assumes 
a universal form which, within a given accuracy, is independent of
redshift and just weakly dependent on cosmology \citep[e.g.][]{ST,SW}.

We measure the linear bias for the haloes in our simulations as follows.
We first determine the functions $b_{\rm hh}$  and $b_{\rm hm}$ 
by directly applying equations (\ref{BHHDEF}) and (\ref{BHMDEF}). 
Within the statistical uncertainties, 
both functions approach asymptotically to a constant on large scales ($k<0.05
\,h$
Mpc$^{-1}$). We use the average of the bias function
measured in the range  $0.01<k<0.05\, h$ Mpc$^{-1}$ (4 $k$-bins)
as our estimate of the linear bias. The standard error of the mean is
used to quantify the corresponding statistical
uncertainty.\footnote{Consistently, in
what follows, we use the rms value as the error on $b(k,M,z,\FNL=0)$.
For $\FNL \neq 0$, we assume that the relative error
is the same as in the Gaussian case.}

\begin{figure}
\includegraphics[width=8cm]{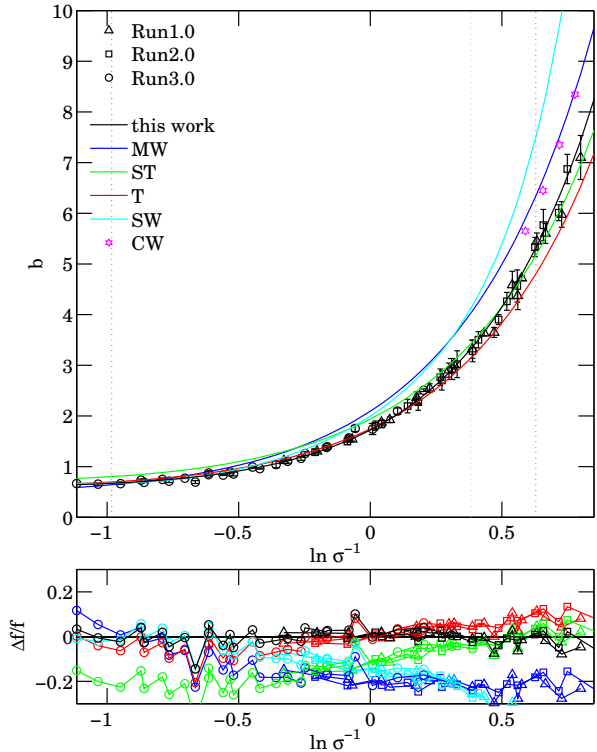}
\caption{Linear halo bias from the Gaussian simulations
Run 1.0 (triangles), Run 2.0 (squares), and Run 3.0 (circles), 
as a function of $\SGI$. 
Only bins containing more than 1000
haloes are shown.
The solid lines correspond to the functions listed in Table 
\ref{TAB_GB} as indicated by the labels.
The four hexagons correspond to the data at $z=10$ by \citet{CW}.
The vertical dotted lines indicate
the maximum and minimum $\SGI$ considered by \citet{TWZZ} (red) and \citet{SW} 
(cyan, in this case the minimum $\SGI$ coincides with the frame
of the figure).}
\label{FIG_BIASGAUSSUNIV}
\end{figure}

In Figure \ref{FIG_BIASGAUSSUNIV} we show the linear
bias obtained from Run 1.0 (triangles), Run 2.0 (squares) and Run 3.0
(circles) as a function of $\SGI$.
Simulation data from snapshots between $z=0$ and $z=2$ are compared
with the commonly used parameterisations listed in Table \ref{TAB_GB}.
Our results are in good agreement with the fit by \citet{SMT} for large masses
and with that by \citet{TWZZ} for smaller masses.
Note that by combining simulation boxes we are able to explore a 
larger interval of $\SGI$ than previous studies.
\begin{table*}
  \caption{Commonly used parameterisations for 
the linear bias arising from Gaussian initial conditions.}
\label{TAB_GB}	 
\begin{tabular}{c|c|c|c|}
\hline
Acronym &
Reference & Functional form & Parameters and Variables \\
\hline
MW &\citet{MOWHITE}    & $b_{\rm MW} = 1 +
\frac{\DELTAC}{\SG^2}-\frac{1}{\DELTAC}$ &
$\DELTAC = 1.686$\\
\\
ST &\citet{SMT}	& $b_{\rm ST} = 1 +\frac{1}{\sqrt{a}\DELTAC}
\left[\sqrt{a}\left(a \frac{\DELTAC^2}{\SG^2}\right) +  \sqrt{a} b \left(a
\frac{\DELTAC^2}{\SG^2}\right)^{1-c} -\right.$ & $\DELTAC=1.686$\\
\\
 & & 
$\left.-\frac{\left(a
\frac{\DELTAC^2}{\SG^2}\right)^c}{\left(a \frac{\DELTAC^2}{\SG^2}\right)^c +
b(1-c)(1-c/2)}\right]$ & $a=0.707, b=0.5, c=0.6$\\
\\
SW & \citet{SW}	& $b_{\rm SW} = 0.53 + 0.39(x^{0.45}) + \frac{0.13}{(40x+1)} + 5
\times 10^{-4}x^{1.5}$ & $x = \frac{M}{M_{\star}}$ with $\sigma(M_*)=1.686$ 
\\
\\
T &\citet{TWZZ}    & $b_{\rm T} = b_{\rm ST}$ & $\DELTAC=1.686, a=0.707, b=0.35,
c=0.80$\\
\hline
\end{tabular}
\end{table*}

Given that no existing model for the linear bias accurately reproduces
our results over the entire mass range spanned by the simulations, 
we decided to derive a new fitting formula. 
In particular, we parameterised the outcome of our simulations as
\BE
b_0=B_0+B_1 \,\SGI+B_2\, \sigma^{-2}\;,
\EE
and used $\chi^2$ minimisation to find
\begin{eqnarray} 
\label{EQ_BESTBIASGAUFIT}
B_0 &= &  0.647 \pm  0.010 \\
\nonumber
B_1 &= & -0.540  \pm 0.028\\
\nonumber
B_2 &= &  1.614  \pm  0.019 \\
\nonumber
\end{eqnarray}
This fit (which reproduces the numerical data with great accuracy
in the range $-1.1<{\rm ln}\, \SGI<0.8$)
should be considered as the linear bias naturally associated with
the mass function given in equations (\ref{EQ_OURFIT}) and
(\ref{EQ_BESTGAUFIT}).

\begin{figure}
\includegraphics[width=8cm]{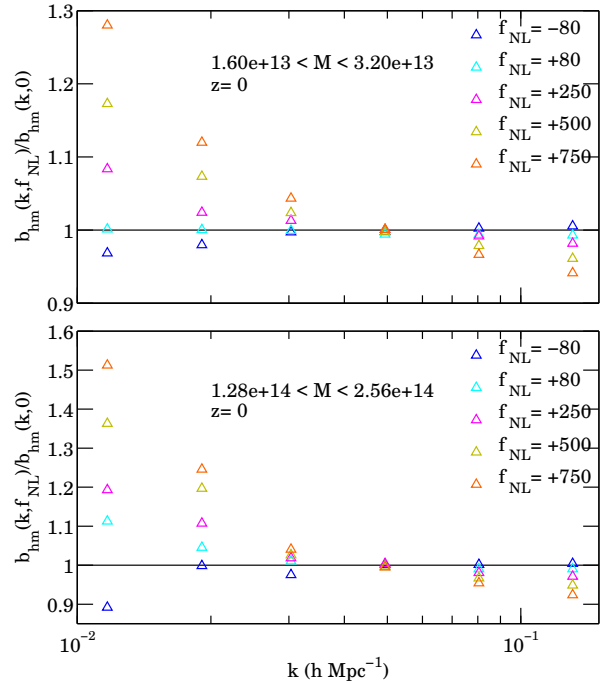}
\caption{Scale-dependent halo bias arising from non-Gaussian
initial conditions. Results are shown in terms of the ratio between the bias
functions 
measured from a simulation with a given $\FNL$ and with $\FNL=0$
at fixed halo mass (indicated by the label in units of $h^{-1} M_\odot$).
Note that in the Gaussian case the bias keeps nearly constant for
$k<0.05\,h$ Mpc$^{-1}$.}
\label{FIG_BIASVSK}
\end{figure}

\subsection{Halo bias from non-Gaussian initial conditions}
\begin{center}
\begin{figure*}
\includegraphics[width=8cm]{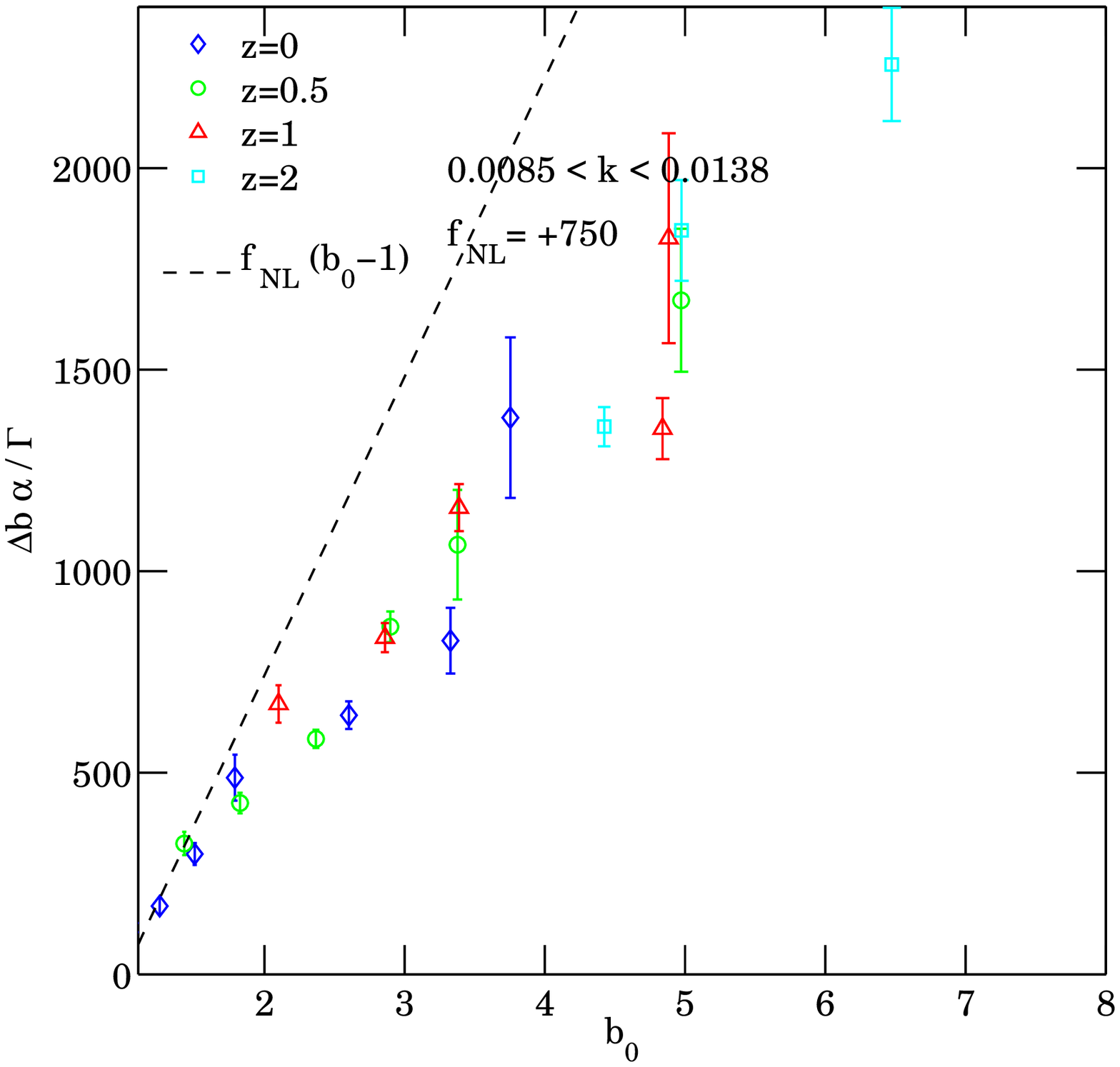}
\includegraphics[width=8cm]{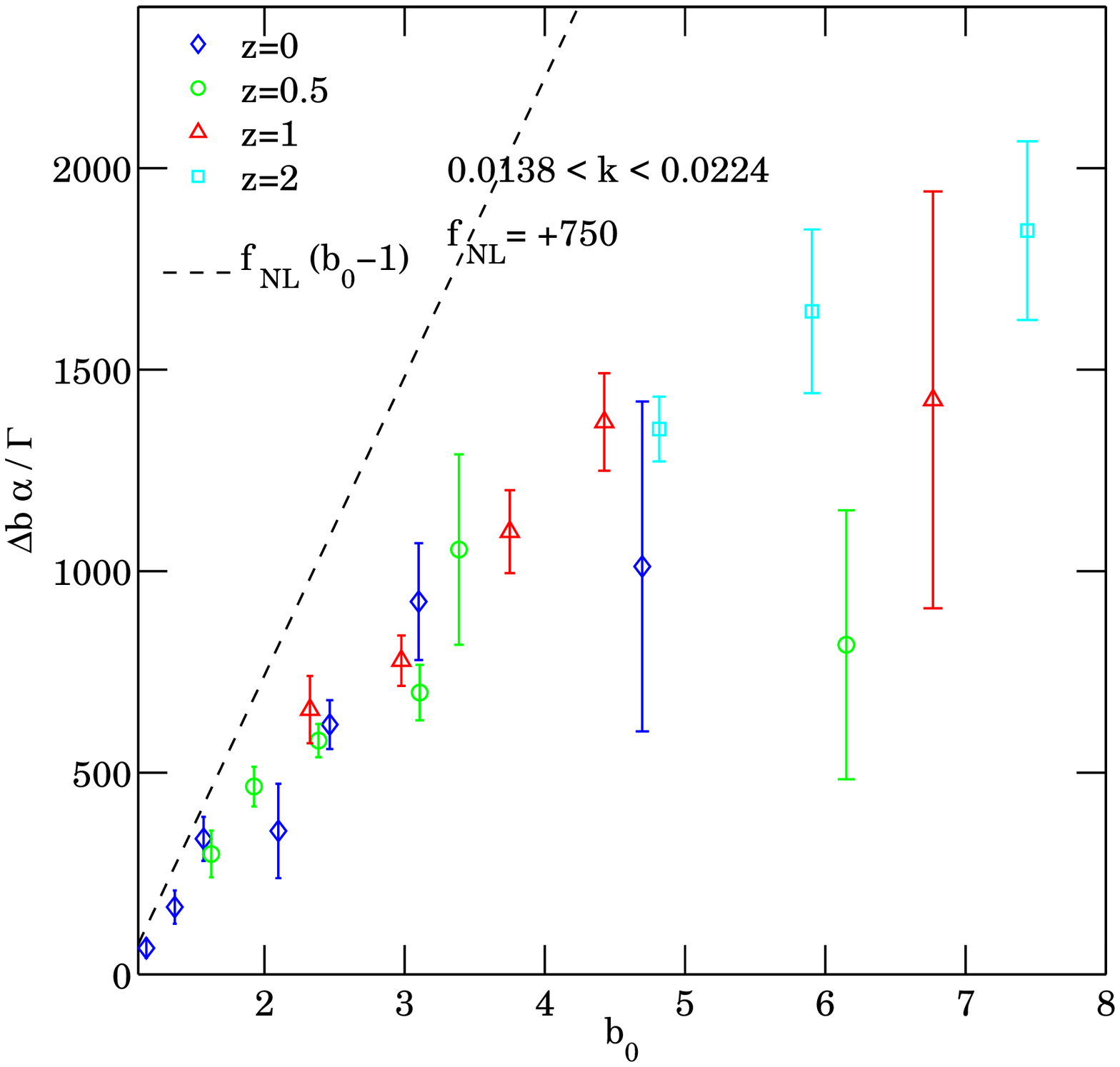}
\includegraphics[width=8cm]{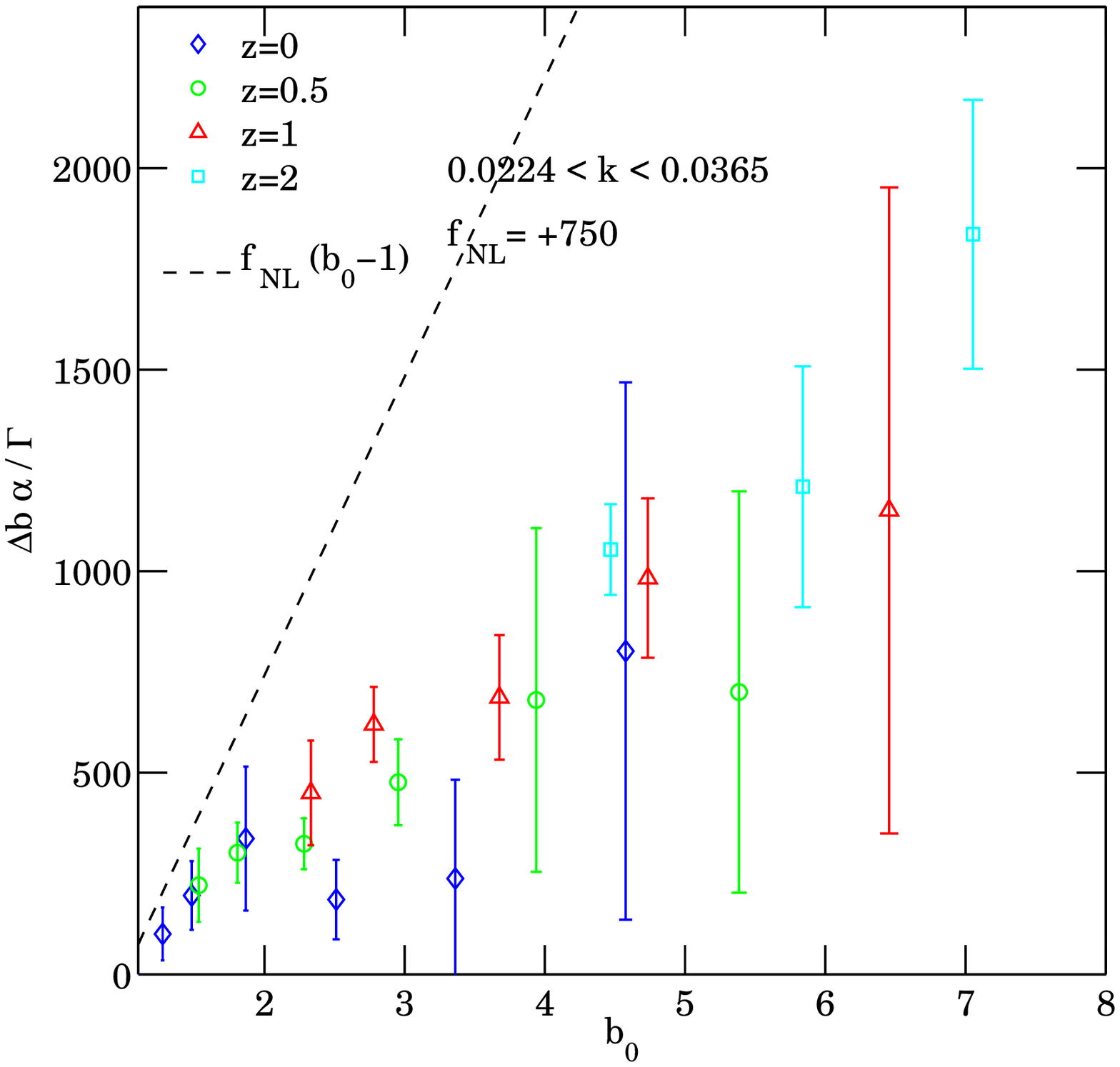}
\includegraphics[width=8cm]{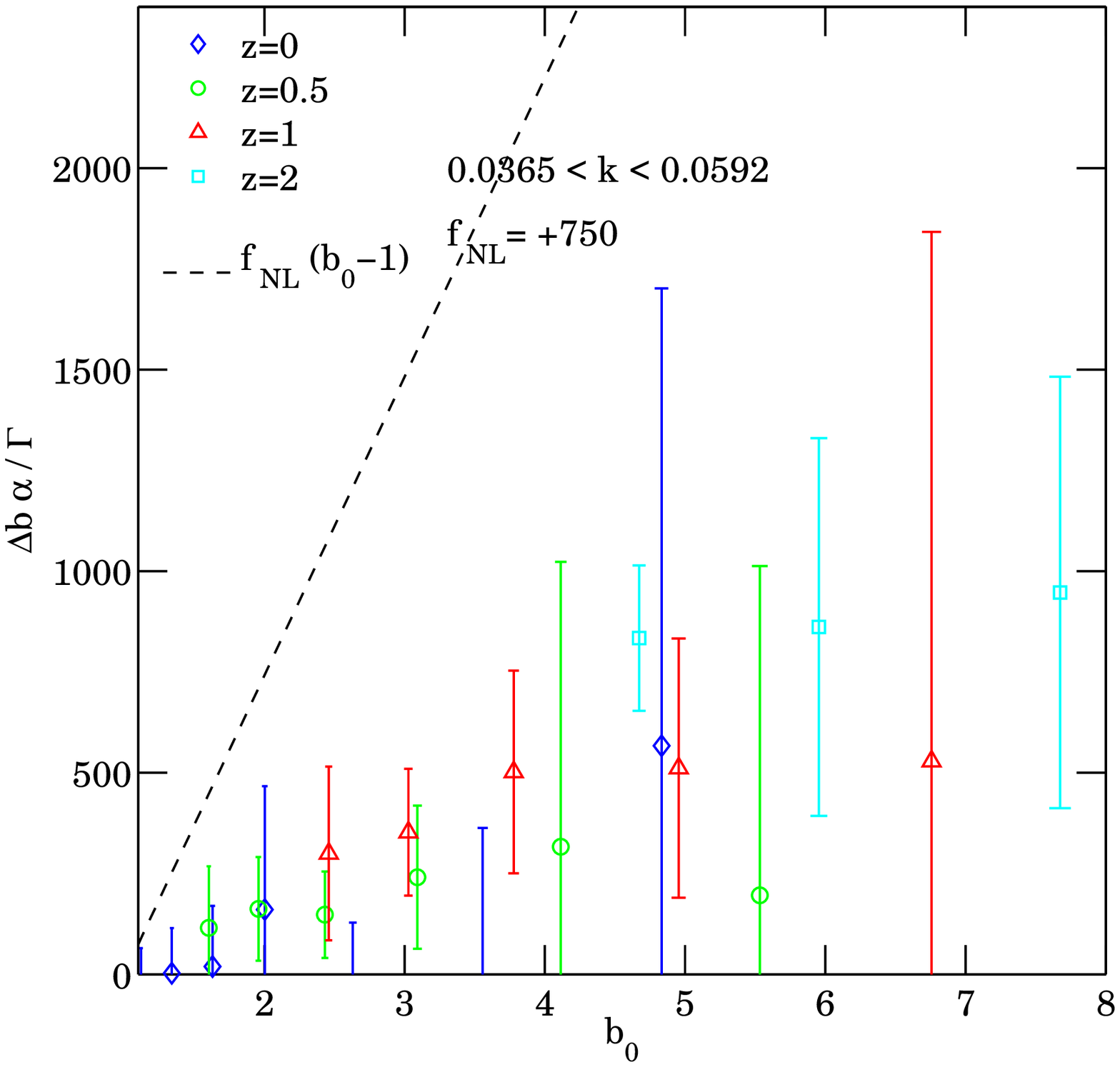}
  \caption{The non-Gaussian bias correction $\Delta b$ as a function
of the linear Gaussian bias $b_0$ for $\FNL=750$. Data from the simulations
are indicated by symbols with errorbars and correspond to different redshifts
as indicated in the label. The dashed line marks the prediction of the
model in equation (\ref{DELTAB}). 
The wavenumber in the labels is given in units of $h$ Mpc$^{-1}$. 
Horizontal errorbars are not drawn to improve readability.}
\label{FIG_NGB1}
\end{figure*}
\end{center}

Recent analytical models, have suggested that the halo bias arising
from non-Gaussian initial conditions of the local type 
does not tend to a constant on large scales.
Rather, the deviation from the Gaussian case should follow
\begin{eqnarray}
\Delta b&=&b(k,M,z,\FNL)-b(k,M,z,\FNL=0)= \nonumber\\ 
&=&
3\,\FNL\,[b_0(M,z)-1]\,\frac{\delta_{\rm c}}{D(z)}\,\frac{g(\infty)}{g(0)}
\,\frac{H_0^2}{c^2}\,
\frac{\Omega_{\rm m}}{k^2\,T(k)}\;,
\label{DELTAB}
\end{eqnarray}
where $\delta_{\rm c}=1.686$, $c/H_0=2997.9\, h^{-1}$ Mpc is the Hubble radius,
$T(k)$ is the matter transfer function,
and $D(z)$ is the linear growth factor of matter perturbations
normalised to unity at $z=0$
\citep{DALAL, MV, SLOSAR, AT, DONALD}.\footnote{The factor
$g(\infty)/g(0)$ is needed since
\cite{DALAL} and \cite{SLOSAR} 
normalise the growth factor $D(z)$ to be $(1+z)^{-1}$ during matter domination.}
The numerical simulations by
\citet{DALAL} have indeed shown that the halo bias is scale dependent
even for small values of $k$
in non-Gaussian cosmologies (with $|\FNL|=100, 500$)
and found qualitative agreement with equation (\ref{DELTAB}).
In Figure \ref{FIG_BIASVSK}, we show how the bias depends on scale
in our simulations which also consider smaller values of $|\FNL|$. 
Our results confirm the presence of a strongly scale-dependent bias.
Larger values of $|\FNL |$ correspond to a more marked
scale dependence. Note, however, that for $k>0.05\,h$ Mpc$^{-1}$
the non-Gaussian deviation $\Delta b$ changes sign. On these scales,
the halo-matter and halo-halo spectra emerging from non-Gaussian perturbations
has actually
less power than in the Gaussian case. The opposite happens with the
matter power spectrum (even to a larger degree) and
the net effect
is a negative $\Delta b$.
This result implies that equation (\ref{DELTAB})
can only hold asymptotically on very large scales.
This is not suprising if interpreted within the peak-background-split
formalism where the large-scale bias is linked to the first derivative 
of the mass function with respect to $\sigma^{-1}$. In the non-Gaussian case
the bias is composed of two parts, a scale-independent term and the correction
given in equation (\ref{DELTAB}).
Since the halo mass function changes shape when 
$\FNL$ is varied, also the constant bias should depend on $\FNL$ for a fixed
halo mass. 
Increasing $\FNL$ corresponds to a larger abundance of massive haloes 
and to a slightly smaller constant bias with respect to the Gaussian case.
Likely, this is what makes the $\Delta b$ in the simulations negative for
positive $\FNL$.
To proceed with a detailed analysis of our simulations,
we find it convenient to rewrite equation (\ref{DELTAB}) as
\BE
\Delta b= \FNL\,(b_0-1)\,\frac{\Gamma}{\alpha(k,z)}\;,
\EE
where $\Gamma=3\,\delta_{\rm c}\,\Omega_{\rm m}\,H_0^2/c^2$
and $\alpha(k,z)=k^2\,T(k)\,D(z)\,g(0)/g(\infty)$.
In Figure \ref{FIG_NGB1}
we test the scaling of $\Delta b$ with redshift, linear bias
and wavenumber for $\FNL=+750$ (where we have the best 
signal-to-noise ratio at high halo mass).
Similar results are obtained with different values of $\FNL$.
The quantity shown is $\Delta b\,\alpha/\Gamma$ which should
correspond to 
$\FNL\,(b_0-1)$ if the analytical model provides a good description
of the data. This quantity is indicated by a dashed line.
The following two trends clearly emerge from the data.
For small values of $k$, the model overestimates the data by ~20-70 per
cent increasing with $b_0$ and independently of $z$.
On smaller scales, discrepancies become more and more severe.
At $k\sim 0.05\ h$ Mpc$^{-1}$, the model is systematically
a factor of 5 higher than the data.
%
The $k$-dependence of $\Delta b$ is therefore different than
in equation (\ref{DELTAB}).

The data also drop a hint that, for $k>0.01\ h$ Mpc$^{-1}$,  
the scaling with $b_0-1$ might only persist up to a maximum value of $b_0$,
$b_{0,{\rm max}}$. For $b_0>b_{0,{\rm max}}$ it appears that the values
of $\Delta b$ are always smaller than expected from the extrapolation of the 
trend $b_0-1$ determined at smaller $b_0$. 
The value of $b_{0,{\rm max}}$ seems to depend both on redshift and
wavenumber and roughly corresponds to constant halo mass for a given $k$.
However, uncertainties in $\Delta b$ at these high masses become very large and 
it is difficult to judge how robust the existence of $b_{0,{\rm max}}$ really
is.
We note anyway that when we tried to fit data at different redshifts (for
a given $\FNL$ and $k>0.015\ h$ Mpc$^{-1}$) 
by adding a variable normalisation constant 
in front of equation (\ref{DELTAB}),
we systematically obtained significantly different fits 
(at a confidence level of 
2.5 $\sigma$) at different redshifts. This trend disappears when only the lowest
values of $b_0$ are considered at each redshift for the fit.   \\

\begin{center}
\begin{figure*}
\includegraphics[width=8cm]{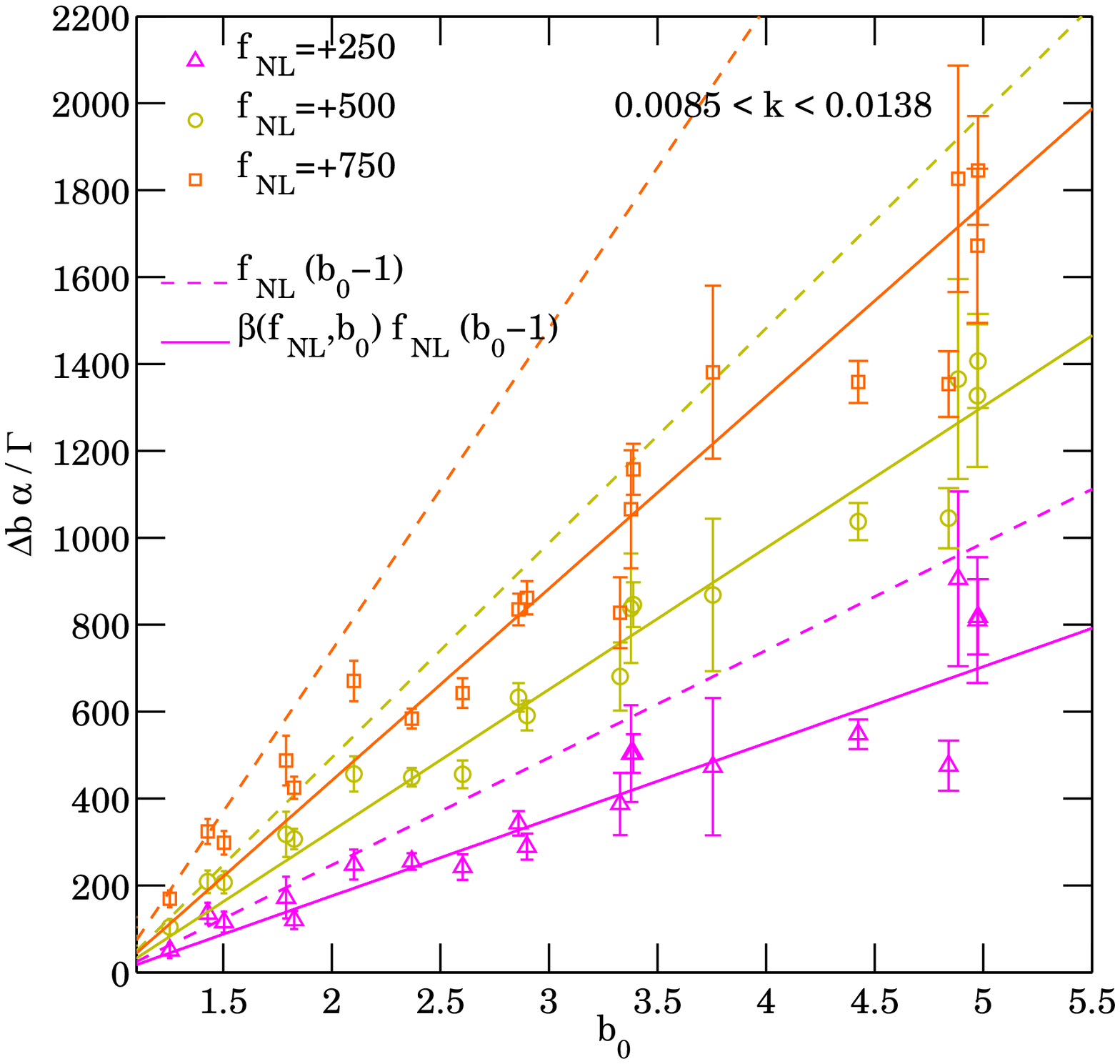}
\includegraphics[width=8cm]{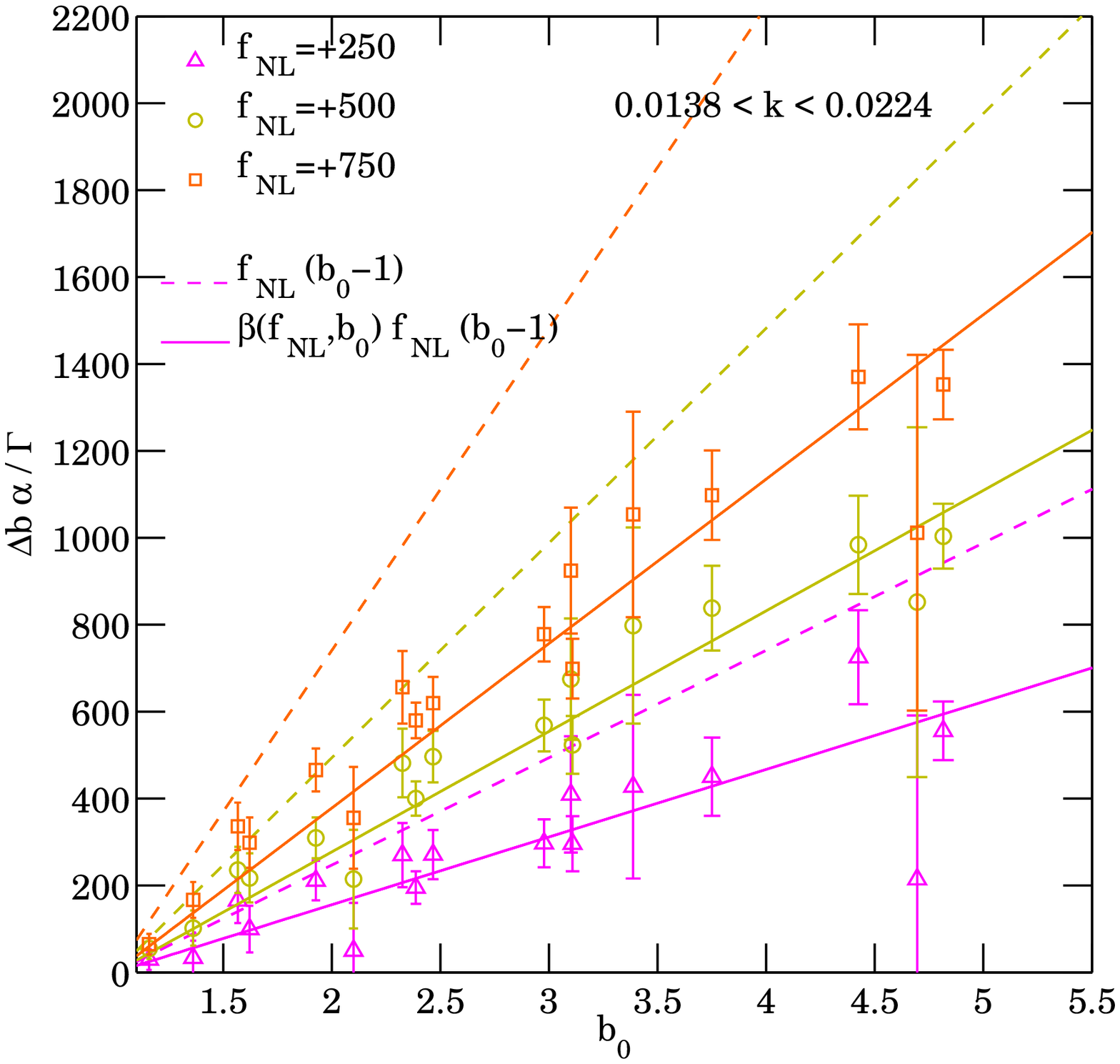}
\includegraphics[width=8cm]{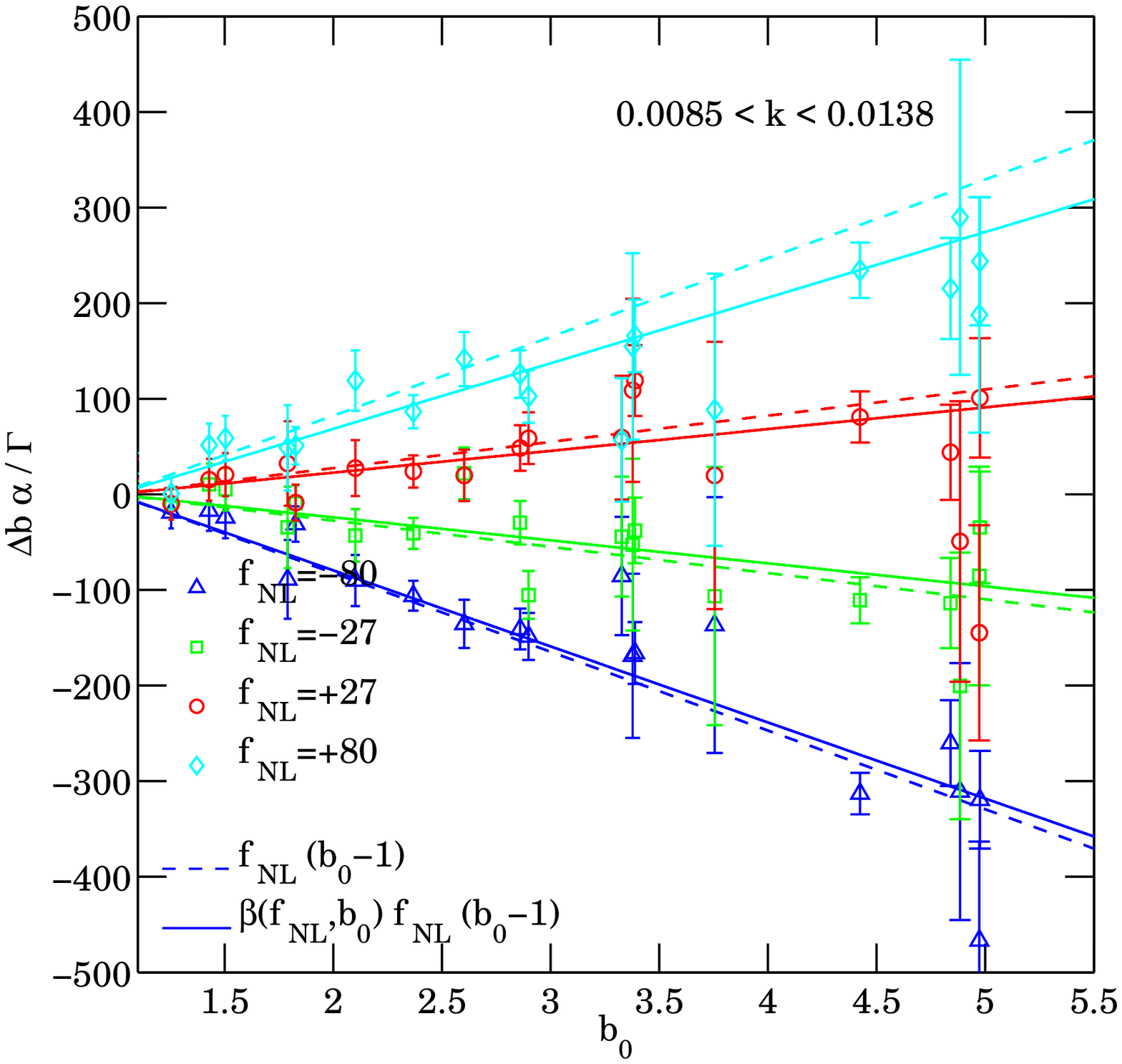}
\includegraphics[width=8cm]{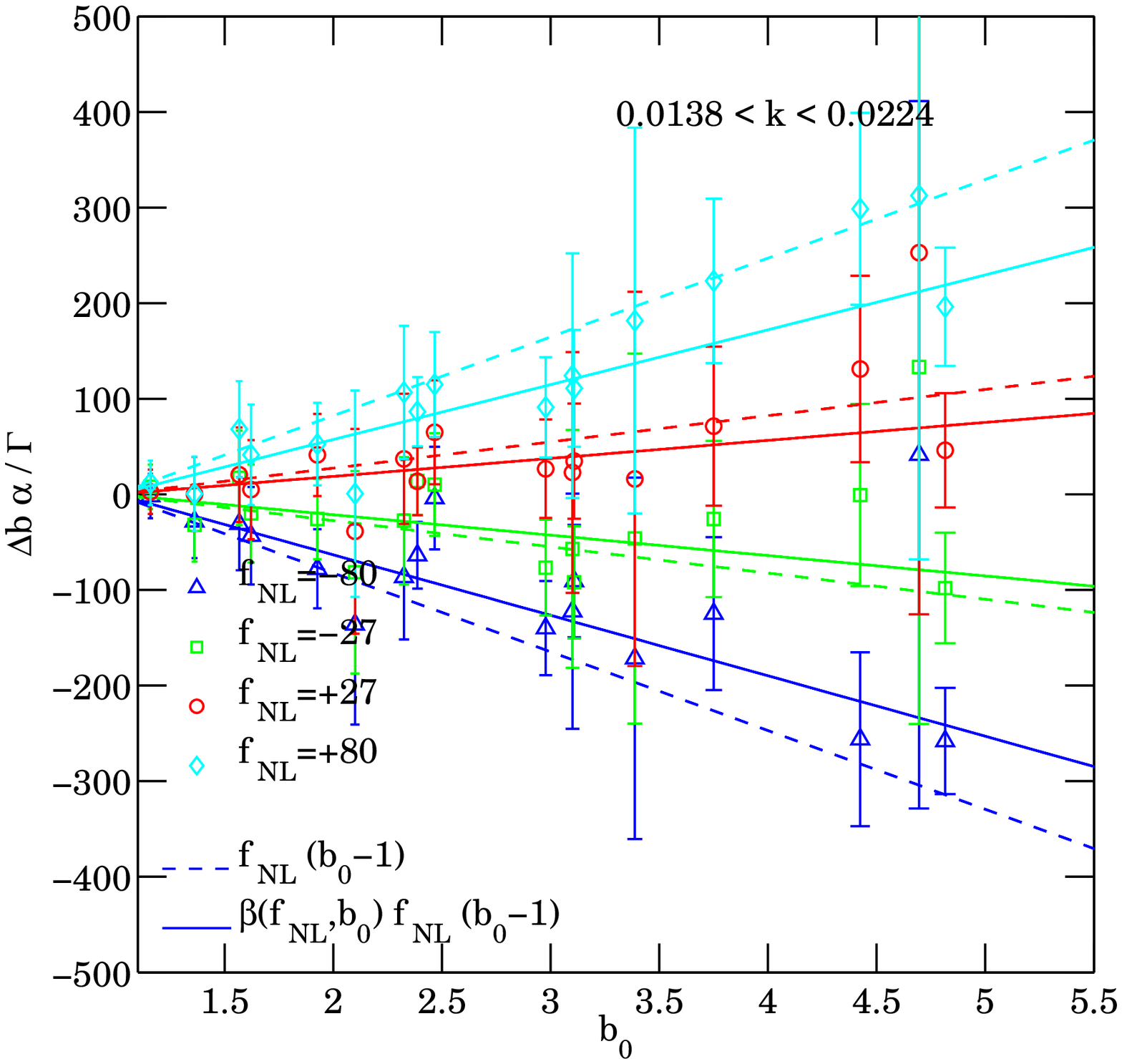}
  \caption{As in Figure \ref{FIG_NGB1} but for all the simulations
of our main series and without distinguishing data from different redshifts. 
The continuous line corresponds to our
best-fitting values of $\beta$ listed in Table \ref{TABBETA}.}
\label{FIG_NGB2}
\end{figure*}
\end{center}

Data from simulations with all the considered
values of $\FNL$ are shown with different symbols and colors
in Figure \ref{FIG_NGB2}.
Each panel refers to a particular wavenumber bin (indicated by the label in
units of $h$ Mpc$^{-1}$). 
The model in equation (\ref{DELTAB}) is again indicated by a dashed line.
Note that,
in most cases, it substantially deviates from the simulation data.
In particular, $\Delta b$ measured from the simulations
shows a much stronger $k$-dependence than the analytical formula,
as already seen in Figure \ref{FIG_NGB1}.
In general,
the overall amplitude of $\Delta b$ drops by an extra 
factor of $\sim 3$ with respect to $k^2\,T(k)$ when moving
from $k\sim 0.01\,h$ Mpc$^{-1}$ to $k\sim 0.05 \,h$ Mpc$^{-1}$
independently of $b_0$ and $\FNL$.
Also, $\Delta b$ does not seem to scale linearly with $\FNL$
while its linear dependence on $b_0-1$ appears to be solid, at least 
for $b_0<b_{0,{\rm max}}$.
We thus introduce a correcting factor $\beta(\FNL,k)$ defined by
\BE
\label{DELTABCORRECTED}
\Delta b= \beta(\FNL,k)\,
\FNL\,(b_0-1)\,\frac{\Gamma}{\alpha(k,z)}
\EE
and we measure it by fitting the simulation data 
for $b(k,M,z,\FNL)$ and $b(k,M,z,0)$
at constant values
of $\FNL$ and $k$. We use an effective variance weighted least squares
method
to simultaneously account for errorbars on both bias parameters.
The best-fitting values are reported in Table \ref{TABBETA} and can
be used to compute the function $\beta$ by interpolation.
The final expression for $\Delta b$, corrected with the $\beta$ factor,
is shown in Figure \ref{FIG_NGB2} with solid lines.

\begin{table*}
\begin{center}
\caption{Best-fitting value and 1$\sigma$ uncertainties for the
multiplicative correction $\beta(k,\FNL)$. The first set of data
corresponds to the $k$-interval where the Gaussian bias is constant.}
\label{TABBETA}
\begin{tabular}{cccccccc}
\hline
$k (h {\rm Mpc}^{-1})$ & $\beta(k,-80)$ &
$\beta(k,-27)$ & $\beta(k,+27)$ &
$\beta(k,+80)$ & $\beta(k,+250)$ &
$\beta(k,+500)$ & $\beta(k,+750)$ \\
\hline
0.0117 &          $0.97 \pm 0.04$ &        $0.88 \pm 0.11$  &       $0.83 \pm
0.13$   &      $0.83 \pm 0.05$   &      $0.71 \pm 0.02$   &     $ 0.66 \pm 0.01$
   &    $ 0.60\pm 0.01$\\
0.0191 &          $0.77 \pm 0.07$ &        $0.78 \pm 0.21$  &       $0.69 \pm
0.25$   &      $0.70 \pm 0.10$   &      $0.63 \pm 0.03$   &     $ 0.56 \pm 0.01$
   &    $ 0.51\pm 0.01$\\
0.0303 &          $0.65 \pm 0.14$ &        $0.58 \pm 0.27$  &       $0.59 \pm
0.27$   &      $0.54 \pm 0.16$   &      $0.50 \pm 0.06$   &     $ 0.42 \pm 0.03$
   &    $ 0.37\pm 0.02$\\
0.0494 &          $0.35 \pm 0.32$ &        $0.45 \pm 0.19$  &       $0.40 \pm
0.26$   &      $0.31 \pm 0.33$   &      $0.29 \pm 0.09$   &     $ 0.25 \pm 0.05$
   &   $ 0.20\pm 0.02$\\
\hline
0.0804 &          $0.01 \pm 0.23$ &        $0.20 \pm 0.21$  &       $-0.10 \pm
0.43$  &      $-0.06 \pm 0.22$  &      $-0.08 \pm 0.18$  &     $-0.10 \pm 0.09$ 
  &    $-0.11\pm 0.04$\\
\hline
\end{tabular}
\end{center}
\end{table*}

Data in Table \ref{TABBETA} have an amazing regularity. 
Apart from a normalisation constant,
each column (row) 
shows the same linear trend with $k$ ($\FNL$). 
This suggests that, within the explored parameter range ($0.01<k<0.05\,h
$ Mpc$^{-1}$ and $-80\leq \FNL \leq 750$), 
\BE
\beta(k,\FNL)=\beta_0\,\left(1-\beta_1\,\FNL \right)\,
\left(1-\beta_2\,k \right)\;.
\label{BETAFIT}
\EE
We thus use this equation to fit the original data for the halo bias from
Gaussian and non-Gaussian initial conditions 
and find
\begin{eqnarray}
\beta_0&=&1.029\pm0.027\;,\nonumber\\
\label{betafit1}
\beta_1&=&(4.25\pm0.33)\times 10^{-4}\;,\\
\beta_2&=&14.8 \pm 0.5\ h^{-1}\,{\rm Mpc}\nonumber\;,
\end{eqnarray}
at the 68.3 per cent confidence level.
Note that we computed the power spectra in finite-sized bins of the
wavenumber, so that 
there is some degree of ambiguity in associating the results with
a given value of $k$. Unfortunately the choice plays a role in determining
$\beta$ as $\alpha$ is a steep function of $k$ on the scales of interest.
In Table \ref{TABBETA} and in equation (\ref{betafit1}), we have used
the arithmetic mean of the wavenumbers contributing to a given bin.
If one instead uses the logarithmic center of the bin, $\beta_0$ is
slightly reduced with a best-fitting value of $0.897 \pm 0.024$.
The parameters $\beta_1$ and $\beta_2$ are unaltered.
Therefore, a systematic contribution $\simeq 0.1$ should be added
to the error budget of $\beta_0$.

equations (\ref{DELTAB}) and (\ref{DELTABCORRECTED}) assume that the
Gaussian bias $b_0$ is constant
with $k$ but this is only approximately true in the simulations.
The fit in equation (\ref{betafit1}), the Table \ref{TABBETA} and the Figures
(\ref{FIG_NGB1}) and (\ref{FIG_NGB2}) have been obtained
by identifying $b_0$ with the actual bias measured in the Gaussian
simulation at each wavenumber. If, instead, the estimate for $b_0$
introduced in Section 5.1 is used, one gets $\beta_0= 0.970 \pm 0.027$,
$\beta_1=(4.13\pm0.33)\times 10^{-4}$
and $\beta_2=13.8 \pm 0.7\ h^{-1}\,{\rm Mpc}$, slightly improving the goodness
of the
fit.

The quadratic dependence of $\Delta b$ on $\FNL$ is rather surprising
as it cannot be straightforwardly derived from the simple models listed above.
It might possibly arise from higher-order terms which have been neglected 
in the expansion that leads to equation (\ref{DELTAB}).
Anyway, it is clearly present in the simulations as it can be seen
by looking at the variation of $\beta$ along a given row in Table
\ref{TABBETA}.
Within the range of $\FNL$ of physical interest, the effect is rather
small: the coefficient $\beta_1$ only corresponds to a few percent
correction. Note that a quadratic
term breaks the symmetry in the amplitude of $\Delta b$ between 
non-linearity parameters with opposite sign and identical absolute value.
It is hard to directly test this against our simulations as we just have
two runs with $\FNL<0$ and both of them correspond to rather small
$|\FNL|$ where the uncertainties in $\beta$ are large.
An alternative explanation for a non-vanishing $\beta_1$
could be that it artificially derives 
from imposing a linear relation in $b_0-1$ to data that do not scale
linearly for $b_0>b_{0, \rm max}$. Indeed, just using datapoints with
small values of $b_0$ we derive bigger values of $\beta$ for large
$\FNL$ (more or less in line with $\beta_1=0$). 
Therefore, what is robust is that at least one of the scalings
with $b_0$ or with $\FNL$ is incorrect in equation (\ref{DELTAB}).
We found that a scaling proportional to $\gamma_0\,(1+\gamma_1\,{\rm log}\,
b_0)$
(with $\gamma_0$ and $\gamma_1$ two adjustable parameters) 
does slightly better (in terms of reduced $\chi^2$) than $\beta_0\,(b_0-1)$,
at least for $k>0.014\,h$ Mpc$^{-1}$. However, since the scaling with
$b_0-1$ has a sound theoretical basis \citep{MOWHITE,CMP} we preferred
to quote our results as in equation (\ref{BETAFIT}).
From the statistical point of view,
the parameters (\ref{betafit1}) provide an acceptable description of the
simulation data to high confidence for all values of $b_0$.
However, they are particularly accurate 
for $b_0>2-2.5$, while $\beta_1\sim 0$ (with the same $\beta_0$ and $\beta_2$)
has to preferred for smaller values of $b_0$.

The linear correction in $k$ should be thought of as the first-order term
of a series expansion in the wavenumber. 
We attempted to determine the corresponding quadratic term by
considering larger values of $k$ in the fit (one bin more, up to $k=0.0962\, h$
Mpc$^{-1}$).
However, values of $\Delta b$ become small compared with the numerical errors
and we found that the quadratic parameter is badly constrained
by the data ($\beta_3=34 \pm 34 \ h^{-2}\,{\rm Mpc}^2$) while the other
parameters remain nearly unchanged (and get larger uncertainties).
Also note that the Gaussian bias starts to depart
from $b_0$ at $k>0.05\ h$ Mpc$^{-1}$ and it is not clear whether 
equation (\ref{DELTAB}) should still be expected to hold in this regime.\\

\citet{DALAL} derived an expression for $\Delta b$ which coincides
with equation (\ref{DELTAB}) but does not include the 
linear transfer function.
Theoretically, this is hard to understand, as
non-Gaussianity is generated well before matter-radiation equality
and one should account for the linear evolution of density perturbations.
Anyway, due to the different $k$-dependence, their expression for $\Delta b$ 
provides a better fit to the simulation data than equation (\ref{DELTAB})
when both models are allowed to vary in amplitude 
with a tunable free parameter.\footnote{The best-fitting value for
this coefficient reads $0.92$ for the model of \citet{DALAL} and $0.58$
for the
model of
\citet{SLOSAR}} None of them, however, provides such an accurate fit to the data
as our
equations (\ref{BETAFIT}) and (\ref{betafit1}), which improve the $\chi^2$ by at
least a factor of $1.7$.

\section{Discussion}
Slosar et al. (2008) have used equation (\ref{DELTAB}) to constrain
$\FNL$ by considering measures of 
the clustering amplitude of luminous red galaxies (LRGs)
and quasars from the Sloan Digital Sky Survey.
Combining all datasets, they found $-29<\FNL<+70$ at 95 per cent confidence. 
How would this result change based on our simulations?
Disentangling the different contributions,
the strongest constraints to $\FNL$ in Slosar et al. (2008)
come from the angular power spectrum of quasars 
with photometric redshifts in the range $1.45<z<2.00$ and a mean bias
of $\sim 2.7$.
Weaker limits are also contributed from the power spectrum of spectroscopic
LRGs and the angular spectrum of photometric LRGs.
(with a bias of $\sim 2$ at $z\sim 0.5$).
Figure (\ref{FIG_NGB2}) and Table \ref{TABBETA} suggest that at the scales
of interest ($0.01<k< 0.05\ h$ Mpc$^{-1}$) the model given in equation
(\ref{DELTAB}) tends to overestimate the scale-dependent bias seen in
the simulations.
Therefore, to match an observed $\Delta b$, 
a larger value of $|\FNL|$ is required than 
predicted by the analytic model. 
When applied to the data by Slosar et al. (2008),
our correction should thus give somewhat looser limits on $\FNL$.
Because of the strong $k$-dependence of the function $\beta$ it is
impossible to give more precise estimates without fitting the power-spectrum
data.
Note, however, that 
a steeper $k$-dependence potentially 
makes determinations of $\FNL$ even more
competitive with respect to studies of CMB anisotropies.

\section{Summary}

We use a series of high-resolution N-body simulations
to study the mass function and the clustering properties of dark-matter haloes 
arising from Gaussian and non-Gaussian initial conditions. 
In particular, we follow the formation of structure in
a universe characterised by the best-fitting parameters from the third- and
fifth-year WMAP data releases.
We consider non-Gaussianity of the local type and 
we use eight different values of $\FNL$
($-80,-27, 0, +27, +80, +250, +500, +750$)
enclosing the parameter space currently allowed by studies of
the cosmic microwave background.
Our main results can be summarised as follows.

(i)
The mass function of dark-matter haloes varies with $\FNL$. Larger
values of the non-linearity parameter correspond to higher abundances
of the most massive haloes.
Analytical models based on the Press-Schechter method \citep{MVJ,LOVERDE}
are compatible with our simulated results for the ratio of the Gaussian and
non-Gaussian 
mass functions only if the critical threshold for halo collapse is 
lowered to $\DELTAC \sim 1.5$.
An accurate fit of the Gaussian $dn/dM$ is necessary to derive
the non-Gaussian mass function from the aforementioned ratio.

(ii)
We find that, in perfect analogy with the Gaussian case \citep{JENK}, 
the halo mass function assumes an approximately universal form.
This means that,
when expressed in terms of suitable variables,
its dependence on redshift and cosmology is
erased to good precision (nearly 10 per cent). 
We parameterise the $\FNL$-dependence of the universal mass function and
provide an accurate fit for its  high-mass end. For $-80\leq\FNL\leq 250$
and for masses $M> 10^{13} \HI M_\odot$, the
best-fitting parameters for the non-Gaussian halo mass function in equation
(\ref{EQ_OURFIT})
are given in equation (\ref{LINEARFIT}) and Table \ref{TAB_MFF_SMALL}. 
This fit reproduces the mass function of friend-of-friends haloes with an
accuracy of 5 per cent on top of a systematic contribution (up to 10
per cent) due to the non perfect universality.
For applications requiring higher precision, an
additional formula is provided: for 
$-80\leq\FNL\leq 80$ and $0\leq z \leq 0.5$ the fit in equations 
(\ref{EQ_NGMFZ}), (\ref{EQ_OURFIT}), and (\ref{EQ_ZANDFNLFIT}) has to be
preferred 
to the universal fit. On the other hand, for higher values 
of $|\FNL|$ and for higher redshifts, the universal fit 
gives a better and more economic (in terms of parameters) description of the 
data.
%
%
In the Gaussian case, we extend the fit to a larger interval of
halo masses ($M>2.4 \times 10^{10} \HI M_\odot$) by combining simulations
with different box sizes: -- see equations (\ref{EQ_OURFIT}) and
(\ref{EQ_BESTGAUFIT}).
Our fitting function provides a precious tool to forecast constraints
on $\FNL$ from future surveys and to analyze current datasets.

(iii)
The matter power-spectrum in non-Gaussian cosmologies departs
from the Gaussian one already on very large scales. 
For values of $\FNL$ within the current CMB constraints
these scale-dependent deviations can be as high
as two per cent at $k=0.3\ h\ {\rm Mpc}^{-1}$ and increase with wavenumber.
The discrepancy is systematic: models with positive $\FNL$ have more large-scale
power than the Gaussian case and models with negative $\FNL$ have less.
This warns against the widespread habit of using the Gaussian matter power
spectrum to determine non-Gaussian bias parameters when high-precision
is required. 
It also suggests that primordial non-Gaussianity modifies the shape and 
amplitude of the
baryonic-oscillation feature in the two-point statistics and the
convergence power spectrum in weak-lensing studies.

(iv)
We present an accurate fitting formula for the linear bias of dark matter haloes
arising from Gaussian initial conditions extending previous work to
larger mass intervals.
This, together with the mass function fit mentioned above, can be used
to constrain parameters of halo-occupation models from clustering data.

(v)
Finally, using 
the halo-matter cross
spectrum, 
we confirm the strong $k$-dependence of the halo bias on large scales
($k<0.05\, h$ Mpc$^{-1}$) which was already detected by \cite{DALAL}. 
However, we show that commonly used
parameterisations based on the peak-background split  
overestimate the effect for $k> 0.01\, h$ Mpc$^{-1}$.
The discrepancy increases with the wavenumber and
at $k > 0.05\, h$ Mpc$^{-1}$
$\Delta b$ in the simulations changes sign with respect to the models.
On top of this,
the analytic model for the scale-dependent part of the bias requires
%
corrections which depend on the non-linearity parameter, the
wavenumber and, possibly, also on redshift and clustering strength. 
equations (\ref{DELTABCORRECTED}) and (\ref{BETAFIT}) with the
best-fitting parameters listed in (\ref{betafit1}) provide a fitting formula
which accurately
reproduces the outcome of the simulations for $0.01<k<0.05\, h$ Mpc$^{-1}$ and
$-80\leq\FNL\leq 750$. 
This 
fit should be employed to constrain $\FNL$ from future clustering
data at low and high redshift. 

\section*{Acknowledgments}
All simulations were performed at the Swiss National Supercomputing Center 
(CSCS)
in Manno, Switzerland. 
AP and OH acknowledge support from the Swiss National Science Foundation.
We thank Volker Springel for kindly making the lean version of the {\sc
Gadget-2} code
available to us.
We acknowledge discussions with Robert E. Smith and Tommaso
Giannantonio.
While our paper was being submitted, 
a related work by Desjacques, Seljak \& Iliev (2008)
appeared as a preprint. Their results based on a 
different halo finder are in qualitative agreement with ours with the
exception of the non-Gaussian bias correction for small halo masses ($b_0<1.5$)
which they find to exceed the model by Slosar et al. (2008).

\appendix
\section{Initial conditions and Zel'dovich transients}

The initial positions and velocities of the particles 
in our N-body simulations have been generated using the Zel'dovich 
approximation.
This method
introduces long-lasting artificial transients in the growth of 
perturbations which might alter the halo mass function at the high-mass end 
even at very late epochs \citep{CROCCE}.
It is therefore important to start the simulation at a sufficiently high
redshift to ensure that all transients have decayed within the cosmic
time at which the simulation output is used for science applications.
Alternatively, less stringent requirements on the initial redshift 
are necessary if one uses second order Lagrangian perturbation theory 
to displace particles at the initial time \citep{CROCCE}.

The simpler Zel'dovich approximation (which only requires the calculation
of the gravitational potential) is much more widespread.
In this case,
a few authors have investigated how to compute the optimal 
starting redshift (see e.g. \citealt{LUK}) as well as quantified
the effects of the initial redshift on the halo mass function 
\citep{TINK} and on the internal properties of dark matter haloes 
\citep{KNEBE}. 
These studies show that simulations of the concordance cosmology (and gaussian initial conditions)
with initial redshifts of $35<z_{\rm start}<60$
(depending on the simulations specifications) have converged
to the correct solution by $z=0$ 
(at least for halo masses $M<10^{14}~ \HI~ \MSUN$).
Even though our $z_{\rm start}$ is in the right ball park,
it is important to test that our results are robust against 
changing initial redshift.
We thus decided to perform the
following simple test:
we re-simulated Run1.0 and Run1.750 of our main Series using
$z_{\rm start}$=99 instead of $z_{\rm start}$=50 and compared the halo mass
functions of the two simulations as a function of redshift (see Figure A1).
In good agreement with \citet{TINK}, for $\FNL=0$ we find that 
discrepancies are smaller than 10 per cent at $z\sim 2$ and that
the (possible) effects of Zel'dovich transients are completely erased by
$z \sim 1$. Our results show that this holds true also for relatively large values
of $\FNL$ (see the right panel in Figure A1 where $\FNL=+750$): even though
non-linearities in the initial conditions are slightly enhanced with
respect to the Gaussian case, at any given redshift the corresponding
density field is also more evolved and the effect of the Zel'dovich
transients on the halo mass function are again erased by $z \sim 2$.
For this reason this paper only uses data from snapshots at $z\le 1.6$
which are accurate to better than 5 per cent.

\begin{center}
\begin{figure*}
\includegraphics[width=8cm]{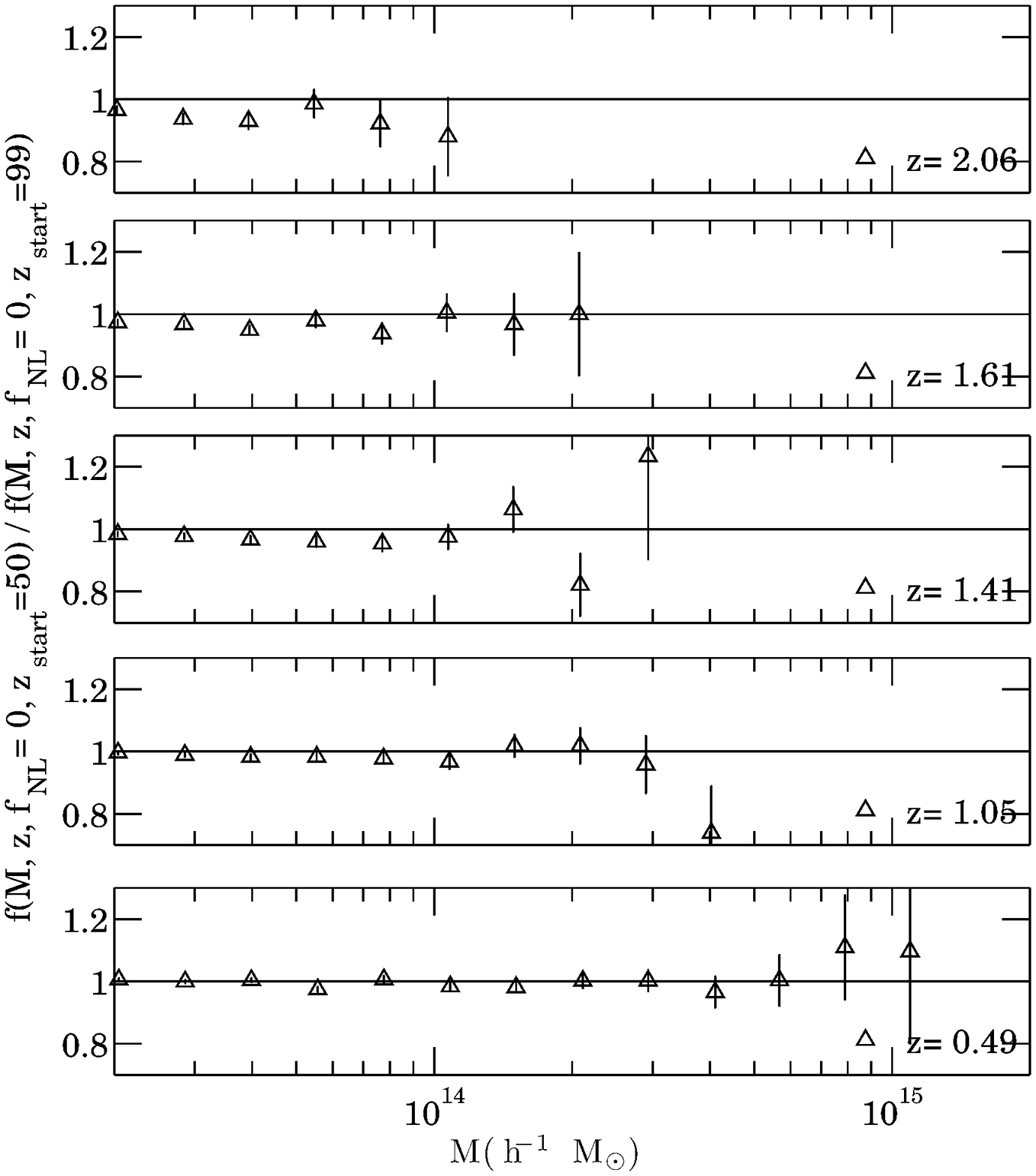}
\includegraphics[width=8cm]{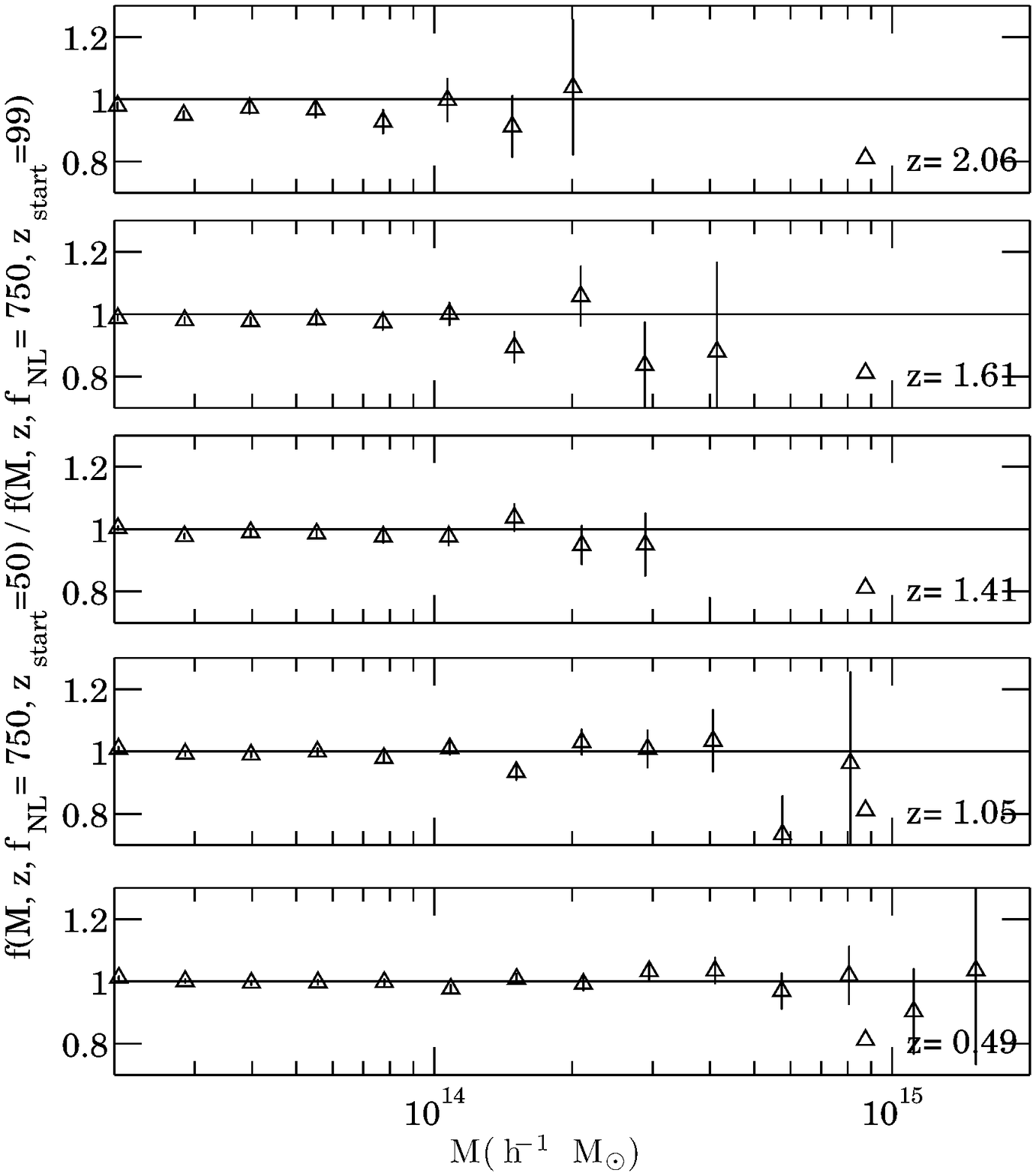}
\caption{Effect of the initial redshift $z_{\rm start}$ on the halo
mass-function, for $\FNL=0$ (left panel) and $\FNL=+750$ (right panel).}
\end{figure*}
\end{center}

\label{lastpage}

\end{document}